\documentclass[twocolumn]{aa}
\usepackage[margin=2cm]{geometry}            
\usepackage{graphicx}
\usepackage{amsmath,amssymb}
\usepackage{xcolor}

\usepackage{epstopdf}
\usepackage{natbib}
\usepackage{comment}

\usepackage{hyperref}
\hypersetup{
    colorlinks=true,
    linkcolor=blue,
    citecolor=blue, 
    urlcolor=cyan,
    }
\usepackage{txfonts}

\newcommand{\dd}{\partial}
\newcommand{\oa}{\overrightarrow}
\newcommand{\lp}{\left(}
\newcommand{\rp}{\right)}

\newcommand{\oav}{\oa{v}}

\newcommand{\oab}{\oa{b}}
\newcommand{\oaB}{\oa{B}}

\newcommand{\oan}{\oa{\nabla}}

\newcommand{\dt}{\delta T}

\begin{document}

\title{Transport of angular momentum and chemical elements by the MRI dynamo in stellar radiative zones}

\author{L. Jouve\inst{\ref{irap1}}, F. Ligni\`eres\inst{\ref{irap1}}, J. Guilet \inst{\ref{cea}}, A. Vanbesien\inst{\ref{imft}}
 }

\institute{Toulouse Univ., Institut de Recherche en Astrophysique et Plan\'etologie, CNRS, CNES, Toulouse, France\\
email:laurene.jouve@irap.omp.eu\label{irap1}
\and D\'epartement d'Astrophysique/AIM 
CEA/IRFU, CNRS/INSU, Univ. Paris-Saclay \& Univ. de Paris Cit\'e 
91191 Gif-sur-Yvette, France \label{cea} 
\and Toulouse Univ., Institut de M\'ecanique des Fluides de Toulouse, CNRS, INP, Toulouse, France \label{imft}}

   \date{Received; accepted }

\abstract{The question of angular momentum transport by magnetic fields has recently been revived by the detection of magnetic fields in the deep interior of red giant stars.}{We aim at characterizing the efficiency of the transport of angular momentum and chemical elements in a stellar radiative zone subject to the magneto-rotational instability, in situations where hydrodynamical instabilities are not triggered.}{We use a large set of zero-net flux shearing-box simulations modelling a portion of a stellar radiative zone located close to the equatorial region. The shear is imposed in the direction of gravity, mimicking a radial differential rotation and rotation is perpendicular to gravity. We aim at establishing scaling laws between the transport efficiency and key parameters such as stratification and rotation.}{We obtain simulations where the MRI is triggered on the self-consistently built longitudinal field and then produces a state of self-sustained turbulence that we associate with dynamo action. We find that in the parameter range explored,
both rotation and stable stratification strongly affect the vertical transport of angular momentum which is always dominated by the Maxwell stress component. In a similar way, we find that the transport of chemical composition is even more strongly affected by the stratification but less severely by rotation. Scaling laws are established and tentative extrapolations to stellar values are discussed.}{Transport by the stratified MRI dynamo is very promising to reconcile stellar evolution models and asteroseismic inversions of internal rotation rates of stars.}


\titlerunning{Transport by the MRI dynamo}
\authorrunning{}
      
\maketitle

\section{Introduction}

Stellar evolution models have made tremendous progress since their early developments decades ago \citep{Kip90}. Since the seminal work of \citet{Zahn92}, they now include the effects of (differential) rotation and associated hydrodynamical instabilities which produce turbulence and a subsequent transport of chemical abundances and angular momentum (AM), with great impact on the whole stellar life \citep[see review by][] {Aerts09}. The inclusion of the transport by magnetic fields is however still an open question. In particular, it is not yet well established how magnetic tension or magnetohydrodynamical instabilities may act to modify the rotation of the various layers of a radiative stellar interior. To progress on such questions, more and more links between 3D numerical simulations tailored at studying the properties of particular hydrodynamical (HD) and magnetohydrodynamical (MHD) processes and 1D stellar evolution models are developed. Examples of such processes are turbulent convection \citep{Trampedach14}, shear-induced turbulence in radiative interiors \citep{Prat13, Prat14, Garaud17}, thermohaline mixing \citep{Traxler11} or transport by magnetic instabilities \citep{Petitdemange23,Meduri24, Barrere25} on which we will come back.

The need for a better description of MHD processes in stellar interiors is now more and more pressing as asteroseismic constraints from exquisite photometric data are established. The example of low mass red giant stars is particularly relevant in this context. Thanks to the 4 years of Kepler data, precise measurements of both internal rotation rates and core magnetic fields are now at our disposal in those stars. These observations come with significant puzzles: some subgiant stars were found to rotate almost rigidly despite the gravitational contraction of their core and the expansion of their envelope \citep{Deheuvels20}, the rotation rate of the core of hundreds of evolved red giant stars were shown to remain constant again despite contraction \citep{Gehan18} and for more evolved subgiants and young red giants, the ratio of envelope to core rotation was found to be moderate, namely only of the order of $\Omega_{core}/\Omega_{env}\approx 10$ \citep{Deheuvels14, Triana17}. Unfortunately, stellar evolution models including state-of-the-art hydrodynamical transport processes are for the moment unable to reproduce the slowly rotating cores of these stars \citep{Ceillier13, Cantiello14}. As stated above, asteroseismology has now also revealed the presence of strong magnetic fields of hundreds of kiloGauss in the core of almost a hundred of red giant stars \citep{Li22, Li23, Deheuvels23, Villate26}. This recent discovery has largely revived the idea that MHD processes could be at play to redistribute the angular momentum in stellar interiors. Beyond the context of red giant stars, angular momentum transport by magnetic fields has been invoked to explain the flat rotation profile of the Sun's radiative interior \citep{Rudi96, Gough98} or the very efficient braking of pre-main sequence stars during their contraction towards the zero-sage main sequence, which also implies AM redistribution within the radiative interior \citep{Gallet13}. A better description of the transport by magnetic fields is thus necessary across all stellar types and evolutionary stages.

Magnetic fields could modify the rotation rate of various regions of stellar radiative zones via different mechanisms. First, magnetic tension is known to counteract the winding of poloidal fields into toroidal fields by the differential rotation. Typical steady-states can then be such that $(B_p\cdot \nabla)\Omega=0$ so that isocontours of the rotation rate coincide with poloidal field lines \citep{Ferraro37}. Additional phase-mixing mechanisms could then act to produce a completely uniform rotation in a very short timescale \citep{Mestel87}. Other MHD processes efficient at redistributing angular momentum are MHD instabilities producing turbulent transport \citep{Spruit99}. The two main instabilities which could develop in the stably stratified environments of stellar radiative zones are the Tayler instability, due to gradients of toroidal fields \citep{Tayler73} and magnetorotational instabilities (MRI) due to velocity shears whose unstable nature is revealed by the presence of weak magnetic fields \citep{Velikhov59, Acheson78, Balbus91}. \citet{Spruit02} proposed a dynamo mechanism relying on both differential rotation and the Tayler Instability and gave a prescription for the associated AM transport. Since then, much interest has been devoted to incorporating these prescriptions in 1D stellar evolution models \citep{Eggenberger19} and improving these prescriptions via analytical developments \citep{Fuller19} or 3D numerical simulations \citep{Zahn07, Petitdemange23, Barrere25}. 
 The transport efficiency of the Tayler-Spruit dynamo proposed by \citet{Fuller19} has been applied with some success in the post-main sequence evolution up to the white dwarf stage but various works tend to show that this mechanism still has difficulties to reproduce the core rotation rates in both the sub-giant and giant phases \citep{Eggenberger19} and that the expected magnetic field produced from this mechanism would be way too low compared to asteroseismic measurements of the core magnetic fields of red giants \citep{Li22}. Those measured strong radial fields could even actually prevent any development of the Tayler instability by their stabilizing effect on toroidal fields \citep{Skoutnev25a}.

The MRI has been somewhat overlooked in the context of stellar radiative zones as strong thermal and compositional gradients are thought to be too stabilizing for turbulence to develop. However, recent works \citep{Jouve20, Gouhier22, Meduri24} have shown that the MRI could actually be favored over the TI in stellar interiors when rotation dominates over the Alfv\'en frequency and that MRI could survive despite stable stratification if large thermal diffusivities were considered or when latitudinal differential rotation was authorized to exist. This could be the case for example of a differential rotation subject to the magnetic tension of a large-scale field which imposes constant rotation along poloidal field lines but differential rotation across them and then possible latitudinal $\Omega$-gradients depending on the large-scale magnetic structure \citep[as in][]{Jouve20, Gouhier22}. The MRI was studied in detail in the context of accretion disks and was found to be a key element for possible dynamo action in a non-convective environment \citep[see][for a recent review on the subject]{Rincon19}. Most of these numerical studies were conducted using the shearing-box framework. Global simulations more adapted to the spherical geometry of stellar interiors then studied the MRI or azimuthal MRI (relying on a dominating toroidal field) but in unstratified environments \citep{Rudi15, Guseva17a} or in setups more relevant to neutron stars \citep{Reboul21}. In any case, very few 3D numerical simulations of stratified MRI were conducted in the context of stellar radiative zones and no clear estimate of the associated transport of AM and chemical abundances were given so far, with the exception of the recent work by \citep{Meduri24} where tentative scalings of the AM and chemical transport was given as a function of the ratio of the Brunt-V\"ais\"al\"a to rotation frequencies $N/\Omega$. Interesting results from this work include a possible weaker dependence on $N/\Omega$ of the MRI dynamo compared to the Tayler-Spruit dynamo and an AM transport stronger than the chemicals transport.
This last property could be particularly appealing for some applications for example to solar-type stars where distinct turbulent coefficients are indeed necessary to reproduce both internal rotation rates and surface Lithium abundances \citep{Dumont21, Eggenberger22}.

In this work, we intend to largely expand this parametric study of stratified MRI for a large range of stratification levels and rotation rates. We establish scaling laws of the turbulent transport as a function of the main parameters relevant for stellar interiors, with the intent to provide prescriptions for new generation 1D stellar evolution models which may help to reconcile models and observations. The equations and numerical model are explained in Sec.\ref{sec_model}, a typical case and the parametric study is detailed in Section \ref{sec_typic}, the effects of stratification, rotation and Reynolds numbers are then respectively discussed in Sections \ref{sec_strat0}, \ref{sec_rot} and \ref{sec_reynolds} and a discussion on the obtained scaling laws and conclusions are given in Sections \ref{sec_discu} and \ref{sec_conclu}.

\section{Model and method}
\label{sec_model}

\subsection{Equations and model geometry}

\begin{figure}[h!]
\centering
   \includegraphics[width=0.5\textwidth]{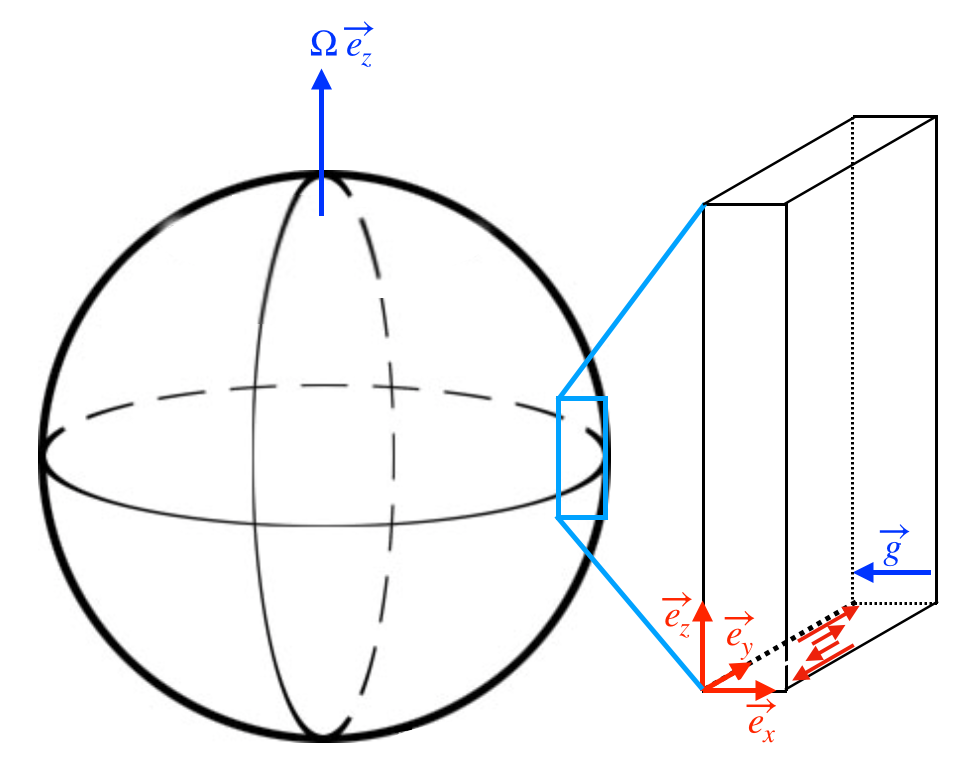} 
   \caption{Schematics of the geometry of the problem: a radial shear localized close to the equator, such that rotation and gravity are perpendicular.}
   \label{fig_schema}
\end{figure}

In this work, our simulations are designed to represent a stably stratified rotating stellar radiative zone, subject to differential rotation (or shear) and possessing initially random magnetic fields. As represented in Fig.\ref{fig_schema}, we wish here to first explore the behaviour of this magnetohydrodynamical problem when the imposed shear is purely radial and where the rotation vector is perpendicular to the direction of gravity. In a star, this would correspond to a shellular (or radial) shear localized in a region close to the equator, as shown in Fig.\ref{fig_schema}. For the simulations, we use the framework of a Cartesian shearing box \citep{Goldreich65} with the coordinates $x$, $y$, and $z$ representing the radial, azimuthal and latitudinal directions respectively. We set the imposed rotation vector $\oa{\Omega}=\Omega \oa{e_z}$ and the linear shear $\oa{V_0}=-S \, x \, \oa{e_y}$, where $S$ is defined as the local value of $-\partial \Omega/\partial \ln r$.

We solve the 3D Boussinesq MHD equations in the aforementioned setup. The magnetic field is expressed in units of a velocity by using a reduced magnetic field $\oab = \oaB/\sqrt{4\pi \rho}$. As a consequence the factor $1/4\pi\rho$ is absent in the momentum equation. In addition, instead of using the temperature variable, we use the buoyancy field  $f=\alpha g \dt$ (with $\dt$ the temperature fluctuation) which has the units of an acceleration. In such a setup, the equations for the solenoidal velocity field $\oav$, the solenoidal magnetic field $\oab$ and the buoyancy field  $f$ read:

\begin{eqnarray}
\centering
  \frac{\dd \oav}{\dd t} + \lp \oav . \oan \rp \oav  &=& - \oan \Pi + f \oa{e_x} + ({S} - 2{\Omega}) v_x \oa{e_y} \\ \nonumber
    &+& {S} x \frac{\dd \oav}{\dd y} + 2 {\Omega} v_y \oa{e_x}  \\ \nonumber
     &+& (\oab . \oan) \oab + {\nu}\Delta \oav 
\end{eqnarray}

\begin{eqnarray}
\centering
	\frac{\dd \oab}{\dd t} + (\oav.\oan) \oab = {S} x \frac{\dd \oab}{\dd y} + (\oab.\oan) \oav  - {S} b_x \oa{e_y} + \eta \Delta \oab
\label{eq_indu_ad}
\end{eqnarray}

\begin{eqnarray}
\centering
   \frac{\dd f}{\dd t} + (\oav.\oan)f  = {S} x \frac{\dd f}{\dd y} - {N}^2 v_x + \kappa \Delta f
\label{eq_temp}
\end{eqnarray}

\noindent where $N$ is the constant Brunt-V{\"a}is{\"a}l{\"a} frequency such that $N^2=\alpha g \,dT/dx$, with $T$ the background temperature and where $\nu$, $\eta$ and $\kappa$ are respectively the constant viscosity, magnetic diffusivity and thermal diffusivity. The total pressure reads:

\begin{equation}
 \Pi = \frac{P'}{\rho} +  \frac{b^2}{2} - \frac{1}{2}(\oa{{\Omega}} \wedge \oa{r})^2 - {S} {\Omega} x^2
\end{equation}

The equations are then adimensioned using ${S}^{-1}$ as the time unit and the box size in the $x$-direction for the characteristic lengthscale, such that $L_x=L$ and $L_y=L_z=6 \, L$ in most calculations. The box is extended in the horizontal directions compared to the radial direction, as shown in Figure \ref{fig_schema}. This choice is motivated by the fact that stable stratification is expected to produce small scales mainly in the direction of gravity $\oa{e_x}$. The reduced box size in the $x$-direction allows us to limit the computational cost while keeping the possibility to have enough structures to perform meaningful spatial averages.

We then define the kinematic and magnetic Reynolds numbers
as $Re ={S} L^2/\nu$ and $Rm = {S} L^2/\eta$. The magnetic and kinetic Prandtl numbers are then respectively defined as $Pm = \nu/\eta=Rm/Re$ and $Pr=\nu/\kappa$.

Since we are also interested in the transport of chemical elements, we solve an equation for a concentration field, which will be considered as a passive scalar and thus does not participate in the background stratification. We 
consider a background concentration characterized by a constant radial 
gradient $C=C_0 + x \;dC/dx$ and solve the advection/diffusion equation governing $c$, the 
concentration departure from this 
background:

\begin{equation}
   \frac{\dd c}{\dd t} + (\oav.\oan)c  = {S} x \frac{\dd c}{\dd y} -v_x\frac{dC}{dx}+ \nu_c \Delta c
    \label{eq_c},
\end{equation}

\noindent where $c$ will be adimensioned using $L dC/dx$. We can then define another Reynolds number $Rec={S}L^2/\nu_c$ characterizing the diffusion of concentration through the coefficient $\nu_c$ taken in all cases to be equal to $\nu$, such that $Rec=Re$ in all calculations.

It is important to note here that there is a possibility for various types of hydrodynamical instabilities in the aforementioned setup, in particular shear instabilities or GSF-type, which have been studied in similar shearing-box calculations for example by \cite{Prat13, Prat14, Garaud15, Barker19, Barker20} and by \cite{Chang21} where both instabilities can coexist. However, for the GSF instability to be excited for radial shear at the equator, the following criterion must be satisfied for the adiabatic and diffusive instabilities respectively \citep{Goldreich67, Fricke68} :
\begin{equation}
2\Omega(2\Omega-S)+N^2 \leq 0 \,\,\, , \,\, 2\Omega(2\Omega-S)+Pr N^2 \leq 0.
\end{equation}

In both cases, a value of $S>2\Omega$ is at least needed for the GSF instability to occur. Consequently, in all our calculations, the value of $S/\Omega$ is chosen to be less than $2$.
In principle, shear instabilities may be present in our setup, at least in weakly stratified cases where $PrN^2/S^2$ is small and thus likely to be below the threshold of stabilization by stratification \citep{Zahn92}. However, the rotation rate $\Omega$ will always be chosen to be of the order or more than $S$, so that shear instabilities are expected to be stabilized in this fastly rotating environment \citep{Chang21}. In any case, we verified that all the simulations computed here are stable with respect to hydrodynamical instabilities and the presence of magnetic fields is consequently necessary for turbulence to develop.

\subsection{Numerical method}

In order to solve the Boussinesq MHD equations presented in the previous section, we use the Cartesian pseudo-spectral code SNOOPY \citep{Lesur05}, which has been previously used for a large number of MRI simulations in various contexts \citep[e.g.][]{Lesur08, Guilet15}. SNOOPY makes use of the so-called shearing box model \citep{Hawley95}. In such a model, the $y$
and $z$ directions are taken as periodic while shear-periodicity is
imposed in the $x$ direction. As the box is shear-periodic, any numerical solution can be decomposed into a set of shearing Fourier modes with time-dependent wavenumbers. Each time-step is evaluated using a 3rd
order explicit Runge-Kutta time-stepping scheme and aliasing errors are eliminated using the $2/3$-rule. Divergence cleaning for the solenoidal velocity and magnetic fields is applied at each timestep using a projector method in the Fourier space.

The initial condition both for the adimensioned velocity and magnetic field consists in the superposition of large-scale modes (such that $k/2\pi<10/L$) of random amplitude below $1$. We ensure that this initial condition results in a zero-net flux magnetic field and that this condition holds to machine precision for the whole duration of our simulations, enabling the study of dynamo action. The buoyancy and concentration fields are initially set to $0$.

For all calculations, a spatial resolution of $(N_x, N_y, N_z)=(64, 192, 384)$ is used. For the most turbulent calculations using the highest Reynolds numbers, convergence tests have been performed with a resolution increased to a maximum value of $(N_x, N_y, N_z)=(128, 256, 512)$.

\section{Typical case, diagnostics and parametric study}
\label{sec_typic}

The aim of this work is to focus on MRI-driven dynamos and the associated turbulent transport. We here first describe a typical velocity and magnetic field evolution in a setup where $S/\Omega=1.5$, $N/S=11$, $Pr=10^{-4}$, $Re=765$ and $Pm=16$. 
Owing to the large value of the thermal diffusivity (and thus the small value of the Prandtl number), this case is only weakly stratified, as will be discussed later.

The effects of rotation compared to the shear are also moderate since the Rossby number $Ro=S/\Omega$ is above $1$. This case is thus close to unstratified zero-net flux MRI simulations performed in the context of accretion disks \cite[e.g.][]{Hawley95, Hawley96} where a Keplerian velocity profile with $Ro=1.5$ is considered. In the following, we first recall the main results obtained in this case. We then give details of our parametric study of the effects of stable stratification and rotation, 
showing in particular that our simulations are performed in a regime where stable stratification is measured by a combination of the Brunt-V\"ais\"al\"a frequency $N$ and the thermal diffusivity $\kappa$. 
Finally, we present the results of a linear stability analysis of the MRI in this regime as it brings some insight on the  effects of stable stratification and rotation.

\subsection{Dynamo action in a weakly stratified Keplerian setup}	
\label{sec:dynamo}

\begin{figure}[h!]
\centering
    \includegraphics[width=0.5 \textwidth]{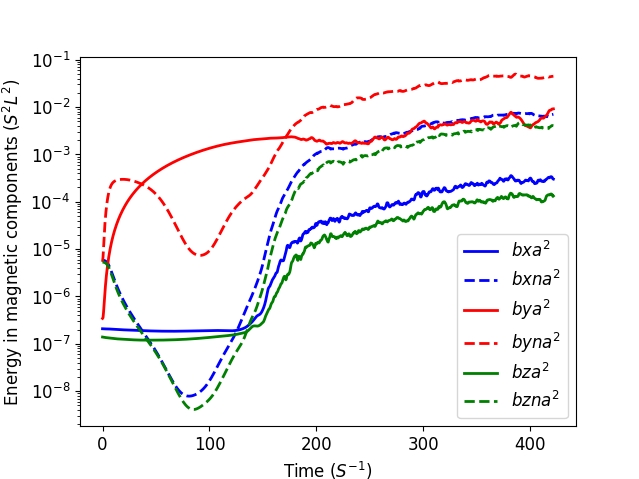}\\
 \vspace{0.5cm}
   \includegraphics[width=0.5 \textwidth]{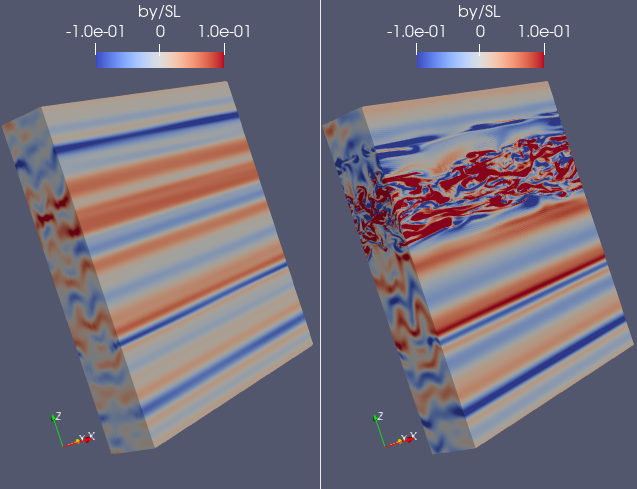}
   \caption{Focus on the initial phase of the instability: top panel: energy in the axisymmetric (denoted by \emph {a}) and non-axisymmetric (denoted by \emph{na}) components of the magnetic field as a function of time for a case with a low level of stratification $Pr=10^{-4}$, $N/S=11$ and a shear to rotation ratio $S/\Omega=1.5$, initiated with a low-amplitude noise below $10^{-3}$. Units are indicated in parenthesis. Bottom panel: snapshot of the $y$-component of the magnetic field adimensioned with $SL$, at $t=100 \, {S}^{-1}$ (left, initiation of the exponential phase) and $t=200 \, {S}^{-1}$ (right, close to the end of the exponential growth).}
   \label{fig_emag_ekin}
\end{figure}

In our fiducial run, the typical evolution is the following: the initial low amplitude magnetic field in the $x$-direction $b_x$ is sheared by the large-scale differential rotation to produce both an axisymmetric and non-axisymmetric longitudinal field $b_y$ which rapidly dominates the magnetic energy. This is shown in the top panel of Figure \ref{fig_emag_ekin}, where the temporal evolution of each component of the magnetic field is presented. We note that to focus on the initial phase here, the initial noise amplitude on the adimensioned magnetic field is set to $10^{-3}$ instead of $1$ as in all the subsequent runs. We quickly get a large-scale axisymmetric longitudinal field concentrated in regions of opposite polarities (to satisfy the zero-net flux condition) as seen on the first snapshot of the bottom panel of Figure \ref{fig_emag_ekin} where the $y$-component of the magnetic field is represented at the beginning of the exponential growth of the instability. The axisymmetric part of $b_y$ then becomes unstable with respect to the magneto-rotational instability, thus producing non-axisymmetric fluctuations of both the velocity and magnetic fields, as seen on the second snapshot of the bottom panel of Figure \ref{fig_emag_ekin}, in which fluctuations grow where the axisymmetric longitudinal field is the strongest. The turbulent fluctuations then spread over the whole box within a typical time $\approx 400 \, S^{-1}$. As we shall describe below, it is then the cross correlations of these non-axisymmetric fluctuations which give rise to a new axisymmetric radial field, ready to be sheared again into an axisymmetric longitudinal field.

As found in all cases computed in this work, a steady-state is reached after a few hundreds of shear timescales, which, for a choice of parameters and specifically with the magnetic Prandtl number $Pm=16$, is sustained for the whole duration of our calculations, namely several thousands of shear timescales. We argue that we are in the presence of a dynamo mechanism, relying on the self-sustained process just mentioned. 

For the purpose of saving computing time, all our cases have usually been calculated for less than a magnetic diffusion time. However, to definitely be able to talk about dynamo action, we need magnetic energy to be sustained at least for a magnetic diffusion time. We thus calculated a run at $Re=382$ and $Pm=4$ up to $2\times 10^4$ shear time scales, translating into a little more than 1 magnetic diffusion time. This run rapidly reaches a steady state which lasts during the whole duration of the calculation. Dynamo action is thus achieved in this case. Most of our runs are then performed at higher $Re$ and higher $Pm$, ensuring that dynamo action is also at play in all our calculations. We note however that stable stratification may displace the critical Reynolds and Prandtl numbers at which dynamo action is triggered compared to this weakly stratified case, as shown for example by \cite{Skoutnev21}. Most of our stratified runs remain turbulent for several thousands of shear timescales in the explored parameter regime, enabling us to calculate the mean transport of angular momentum and chemicals but we note that dynamo action may be lost in some of these more stratified cases if computed on longer timescales.

\begin{figure}[h!]
\centering
   \includegraphics[width=0.5\textwidth]{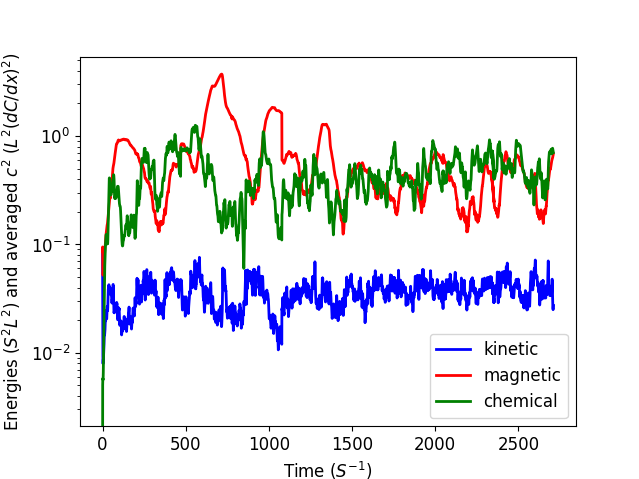}
   \includegraphics[width=0.5\textwidth]{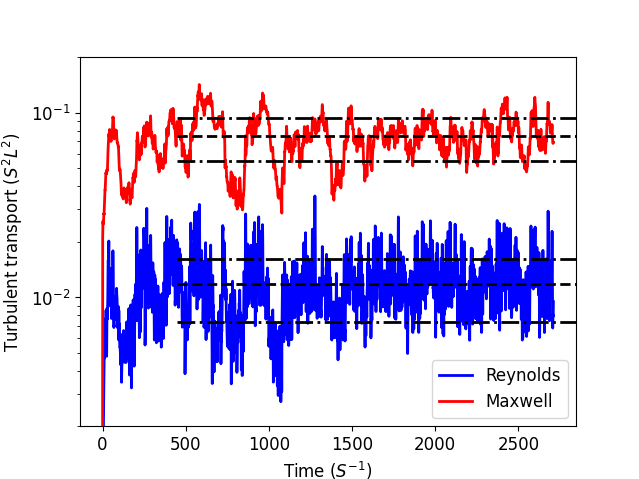}
   \caption{Temporal evolution of the magnetic and kinetic energies and volume-averaged chemical concentration (top) and absolute value of the Maxwell $\langle b_x b_y \rangle$ and Reynolds stresses $\langle v_x v_y\rangle $ as a function of time (bottom) for a case with a low level of stratification $Pr=10^{-4}$, ${N/S}=11$ and a shear to rotation ratio $S/\Omega=1.5$. Units are indicated in parenthesis. On the bottom panel, the superimposed dashed lines indicate the mean value of the stresses calculated from time $t=500 \, {S}^{-1}$ and the dash-dotted lines indicate the 1$\sigma$ deviation to the mean value.}
   \label{fig_emag1e-4}
\end{figure}

The top panel of Figure \ref{fig_emag1e-4} shows the evolution over more than $2700$ shear timescales of the magnetic and kinetic energy densities and of the volume-averaged squared chemical concentration as a function of time respectively defined as:
\begin{equation}
E_{mag}(t)=\frac{1}{2}\int_V \, (b_x^2+b_y^2+b_z^2) \, dV 
\end{equation}
\begin{equation}
E_{kin}(t)=\frac{1}{2}\int_V \, (v_x^2+v_y^2+v_z^2) \, dV 
\end{equation}
\begin{equation}
E_{c}(t)=\frac{1}{2}\int_V \, c^2 \, dV
\end{equation}
The steady state is characterised by a magnetic energy typically higher by one order of magnitude than the kinetic energy, but this ratio is strongly sensitive both to stratification and rotation, as we shall see below. Another known feature that we also find in the weakly stratified case is the cyclic behaviour of the axisymmetric magnetic field. These cycles in MRI-driven dynamos were found before in unstratified simulations in the context of accretion disks \citep[see for example][]{Lesur08}.

We here quickly summarize the self-sustained process at play in this simulation and which was described before for the unstratified sub-critical MRI dynamo \citep[see review by][]{Rincon19}: a configuration dominated by an axisymmetric longitudinal field becomes unstable with respect to the magneto-rotational instability. Fluctuations both in the velocity and magnetic fields then result in an electromotive force producing an axisymmetric radial field which is then sheared again to reproduce a new axisymmetric longitudinal field, effectively closing the dynamo loop. To be more precise, we can write the $xy$-averaged ideal mean-field induction equation, i.e. involving the axisymmetric magnetic field $\overline{\overrightarrow{b}}$ and velocity field $\overline{\overrightarrow{v}}$ and the non-axisymmetric parts $\overrightarrow{b}^{\prime}$ and $\overrightarrow{v}^{\prime}$. We then get the following equations:

 \begin{equation}
\partial_t \overline{b}_x=-\partial_z \overline{\cal E}_y - \overline{v}_z\partial_z\overline{b}_x + \overline{b}_z \partial_z\overline{v}_x
 \end{equation}
 \begin{equation}
\partial_t \overline{b}_y=-{S}\overline{b}_x + \partial_z \overline{\cal E}_x - \overline{v}_z\partial_z\overline{b}_y + \overline{b}_z \partial_z\overline{v}_y
 \end{equation}
where the last two terms in both equations correspond to advection and stretching respectively and have a negligible contribution to the equations, and where the axisymmetric part of the electromotive force involves correlations of the non-axisymmetric parts of $b$ and $v$: $\overline{\cal E}_y=\overline {v^{\prime}_z b^{\prime}_x}-\overline{v^{\prime}_x b^{\prime}_z}$ and $\overline{\cal E}_x=\overline{v^{\prime}_y b^{\prime}_z}-\overline{v^{\prime}_z b^{\prime}_y}$.  
The regeneration of the axisymmetric field in the $x$-direction is then due to the $z$-gradient of the axisymmetric electromotive force in the $y$-direction and thus implies a non-linear process, while the axisymmetric longitudinal field $\overline{b}_y$ is mainly produced by the shearing of $\overline{b}_x$ via the so-called $\Omega$-effect, i.e. a linear process. The growth of the linear MRI is also a necessary ingredient of the dynamo mechanism. The term involving $\overline{\cal E}_x$ was shown to be a sink term rather than a source for the axisymmetric field \citep[e.g.][]{Lesur08a,Mondal26}. It is thus a combination of linear and non-linear processes which guarantees the sustenance of the described dynamo mechanism. Close to the MRI dynamo transition (at small Reynolds numbers), it was shown that the self-sustained process gives rise to a small number of pairs of invariant non-linear solutions that could be either traveling waves \citep{Rincon07} or non-linear cycles \citep{Herault11}. This restricted number of solutions enables to model the electromotive force and thus the whole dynamo process. However, as the Reynolds number increases, any initial condition can lead to fully-developed MHD turbulence and the modeling of the dynamo process using a small number of nonlinear solutions becomes impossible. This constitutes a motivation to explore this mechanism far from the transition where the turbulent cascade can strongly affect the self-sustained process, especially when it is subject to both stable stratification and strong rotation, a regime relevant to stellar interiors. Finally, we note that since our goal is primarily to quantify the efficiency of the MRI dynamo to transport angular momentum and chemical elements, understanding the sustenance of the axisymmetric field components in the fully turbulent dynamo process is not enough since it is in fact the non-axisymmetric parts of the magnetic field components which dominate the transport, as previously reported in unstratified MRI dynamo simulations.

\subsection{ Angular momentum and chemicals  transport} 

It is important to note here that despite the cyclic behaviour of the magnetic energy and thus significant variations during the steady state, the diagnostics for angular momentum transport are still clearly defined, again as previously noted by \cite{Lesur08}. In the context of stellar radiative zones, we are particularly interested in the transport of angular momentum and chemicals in the radial direction, i.e. in our case in the direction of the mean shear. 
The dominant components of the volume-averaged Reynolds and Maxwell stresses in the $x$-direction, namely 
\begin{equation}
\langle v_x v_y\rangle =\int_V v_x v_y \, dV \,\, , \, \langle -b_x b_y\rangle =\int_V -b_x b_y \, dV \,\,
\end{equation}
are shown on the bottom panel of Figure \ref{fig_emag1e-4}. For clarity, the chemical transport $\langle cv_x\rangle $ is not shown as the curve is almost superimposed on the Maxwell stress. To compute the mean stresses, we discard the initial transient phase (which typically lasts around $400-500$ shear timescales) and focus on the steady-state for several hundreds of shear timescales. The mean and standard deviation values are overplotted in black lines on Fig. \ref{fig_emag1e-4} for the Maxwell and Reynolds stresses.

\subsection{Parametric study in the large thermal diffusivity regime}

The main questions around the AM and chemical element transport in stellar radiative interiors are how its properties are impacted by the main ingredients found in those environments, namely stable stratification and rotation. We thus focus in this work on varying the rotation rate $\Omega$ and the stratification in entropy.

The shear rate is kept fixed and serves as the inverse of the unit timescale in all simulations.

To ensure dynamo action, as stated above, a sufficiently high value of the Reynolds number $Re=765$ and Prandtl number $Pm=16$ are used in the first cases and then their impact on the results are tested in section \ref{sec_reynolds}.

Finally, we fix $Rec=Re$ in all our calculations.

To study the effect of the Coriolis force, we will vary the rotation rate $\Omega$ such that the Rossby number 
is varied between $0.2$ and $1.8$. The impact of rotation is analysed in section \ref{sec_rot}.

Without the help of thermal diffusion, the large 
ratio between the  Brunt-V\"ais\"al\"a frequency and the shear rate typical of stellar radiative zones (denoted by $N/S$) would imply both hydrodynamical and MHD stability.
The large thermal diffusion of stellar interiors can mitigate the effects of stable stratification
as thermal diffusion reduces temperature fluctuations and thus the amplitude of the buoyancy force. This process is efficient for small enough length
scales such that the thermal diffusion time
$1/(k^2 \kappa)$ is shorter than the buoyancy time scale $1/N$, where $k$ is the wavenumber associated with the lengthscale. In other words, this also means that we expect only scales at a P\'eclet number $Pe=SL^2/\kappa<1$ to be unstable in our regime where $N/S>1$ and to participate in the transport of angular momentum and chemical elements.

In our current study, we can be more precise on how the thermal diffusivity and the Brunt-V\"ais\"al\"a frequency will combine to measure the effect of stratification. 
The analysis of the different terms of the temperature equation \ref{eq_temp} in a typical simulation indeed shows that it reduces to a balance between the advection against the mean temperature gradient $N^2 v_x$ and the dissipation of temperature fluctuations $\kappa \Delta f$. An example of the calculations of each term in such a typical simulation is shown in appendix \ref{appendix_temp}, where the balance between $N^2 v_x$ and $\kappa \Delta f$ is clearly dominant. 
This holds true for all times in the simulations and for all stratified simulations performed in this study. As a consequence, we can rescale the temperature fluctuations $\delta T$ as $\kappa/N^2 \times \delta T$ and the effect of stratification in the set of equations will be controlled solely by the parameter $N^2/\kappa$ or in dimensionless form by the parameter $PrN^2/S^2$. 
We indeed verified that in all stratified calculations performed in this work, the magnetic and velocity fields, as well as the Reynolds and Maxwell stresses are statistically the same (variations being due to different random initial conditions) for the same value of $PrN^2/S^2$, produced with different pairs of $Pr$ and $N/S$. In the simulations presented here, the value of $PrN^2/S^2$ is varied from $0$ in unstratified cases to $2.5$ in the most strongly stratified cases. The impact of stratification measured by this parameter is studied in Sec. \ref{sec_strat0}.

In the considered small-Péclet-number regime \citep{Lignieres99,Garaud21}, the buoyancy timescale is measured by $t_{BM}=t_B^2/t_\kappa$ where $t_B$ is the buoyancy timescale in a non-diffusive atmosphere and $t_\kappa$ is the thermal diffusion timescale. If, for $t_B$, we take the inverse of the gravity wave frequency $\omega=(k_h/k) N$ where $k_h=\sqrt{k_y^2+k_z^2}$ is the horizontal wavenumber, and for $t_\kappa$ the inverse of the thermal damping rate $1/(\kappa k^2)$, we then get the modified buoyancy timescale $t_{BM}=(\kappa k^2/N^2) \times (k^2/k_h^2)$. In the case where $k_z >> (k_x,k_y)$, we note that the expression simplifies to $t_{BM}=\kappa k_z^2/N^2$.

As the MRI is one of the linear ingredient playing a key role in the self-sustained process, we performed a linear analysis of the MRI of a given axisymmetric longitudinal field $B^0_y$, in the high thermal diffusivity regime. The details of this analysis are shown in appendix \ref{sec_linear}. We can summarize this analysis by stating that the growth rate and Alfv\'en frequency $\omega_A=k_y B_y^0$ of the most unstable mode in the relevant regime (i.e. $(k_x, k_z)>> k_y$) are given by:

 \begin{equation}
  \sigma =\frac{S}{2} \frac{1}{\frac{N^2}{\kappa k_z^2} \frac{1}{4 \Omega}  + 1}
  \label{eq_growth}
  \end{equation}
   \begin{equation}
  \omega_A^2 = \Omega S \frac{1}{\frac{N^2}{\kappa k_z^2} \frac{1}{4 \Omega} + 1} \left(1 -\frac{S}{4\Omega}\frac{1}{\frac{N^2}{\kappa  k_z^2} \frac{1}{4 \Omega} + 1} \right)
  \label{eq_omegaA}
\end{equation}

We see that for the unstratified case $N=0$, Equations \ref{eq_growth} and \ref{eq_omegaA} simplify to $\sigma=S/2$ and $\omega_A^2=\Omega S \left(1 -\frac{S}{4\Omega}\right)$, so that rotation does not impact the MRI growth rate but does increase the Alfv\'en frequency of the most unstable mode. When stratification increases, the growth rate and Alfv\'en frequency start to be affected by both stratification and rotation. Both effects combine in the ratio 
$\frac{N^2}{\kappa k_z^2} \frac{1}{4 \Omega}$, which is proportional to the timescale ratio $\frac{t_\Omega}{t_{BM}}$ where we define $t_\Omega=1/(2\Omega)$. At a fixed level of stratification and fixed shear rate, we then note that increasing the rotation rate favors the instability as it 
increases both the growth rate and the Alfv\'en frequency. 

We finally see in this linear analysis that the wavenumber $k_z$ is not fixed so that the instability may adapt its lengthscale to best accommodate for the effects of stratification and rotation. In particular, a smaller lengthscale in the stratified case will increase both the growth rate and the Alfv\'en frequency of the most unstable mode. However, if $k_z$ becomes too large, dissipative effects may act to suppress the instability. By equating the maximum growth rate to the viscous damping rate (the fastest dissipative process in our high $Pm$ cases), the linear analysis predicts that the system should be stabilized for:
\begin{equation}
\label{eq_limit}
Pr \frac{N^2}{S^2} \ge \frac{2}{3} \frac{\Omega}{S} 
\end{equation}

Those various results of the linear analysis may guide us towards understanding some aspects of the non-linear simulations which will now be presented.

\section{Impact of stable stratification}
\label{sec_strat0}

We start from the fiducial run described above and increase the level of stratification, keeping the same value of the Rossby number
$Ro=1.5$. The value of $PrN^2/S^2$ controlling the stratification is varied from $2.5\times10^{-3}$ to $2.5$, by varying $Pr$ while keeping $N^2/S^2$ fixed. We first describe the flow and field characteristics before discussing the behaviour of the transport of angular momentum and chemicals.

\subsection{On the flow and field characteristics}

\begin{figure*}[h!]
\centering
  \includegraphics[width=\textwidth]{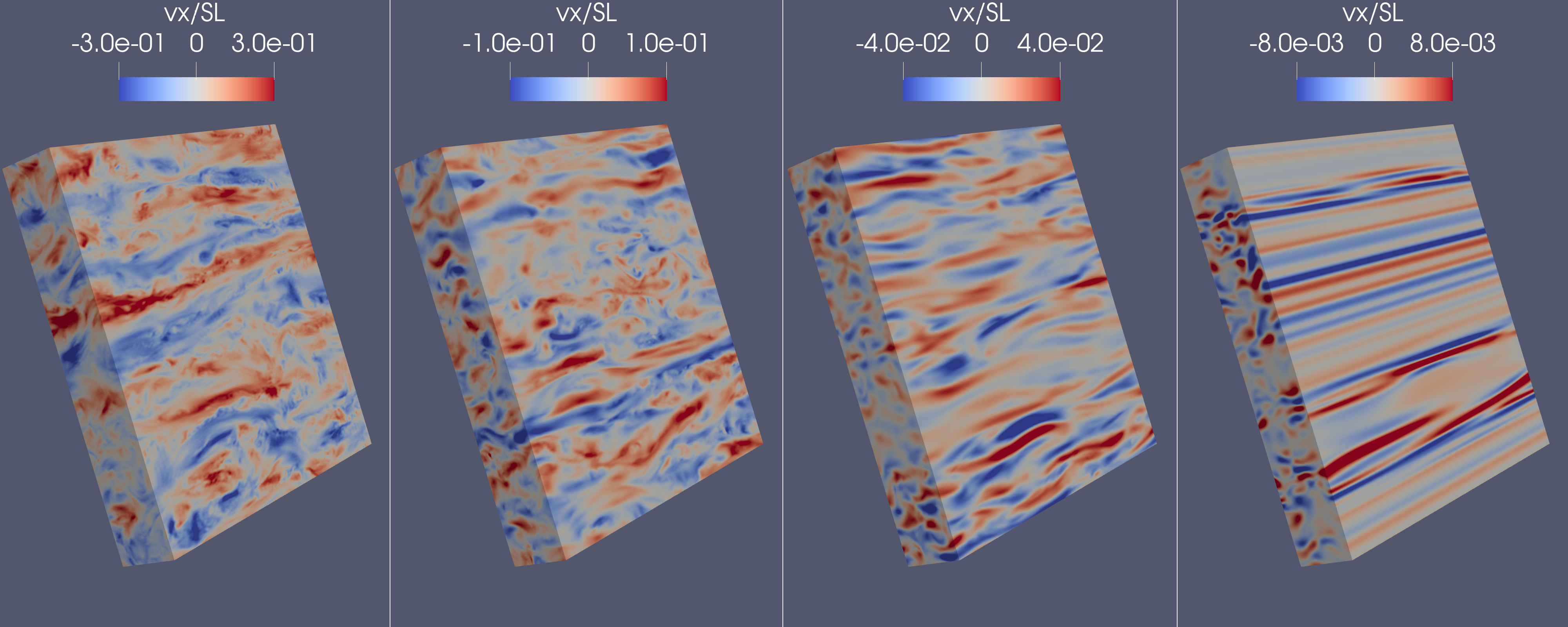} 
   \caption{Effect of stratification on the radial velocity $v_x$, adimensioned with $SL$. From left to right, the values of $PrN^2/S^2$ are $1.2\times10^{-2}, 1.2\times10^{-1}, 4.9\times10^{-1}$ and $2.5$.}
   \label{fig_vx_strat}
\end{figure*}

\begin{figure*}[h!]
\centering
\includegraphics[width=\textwidth]{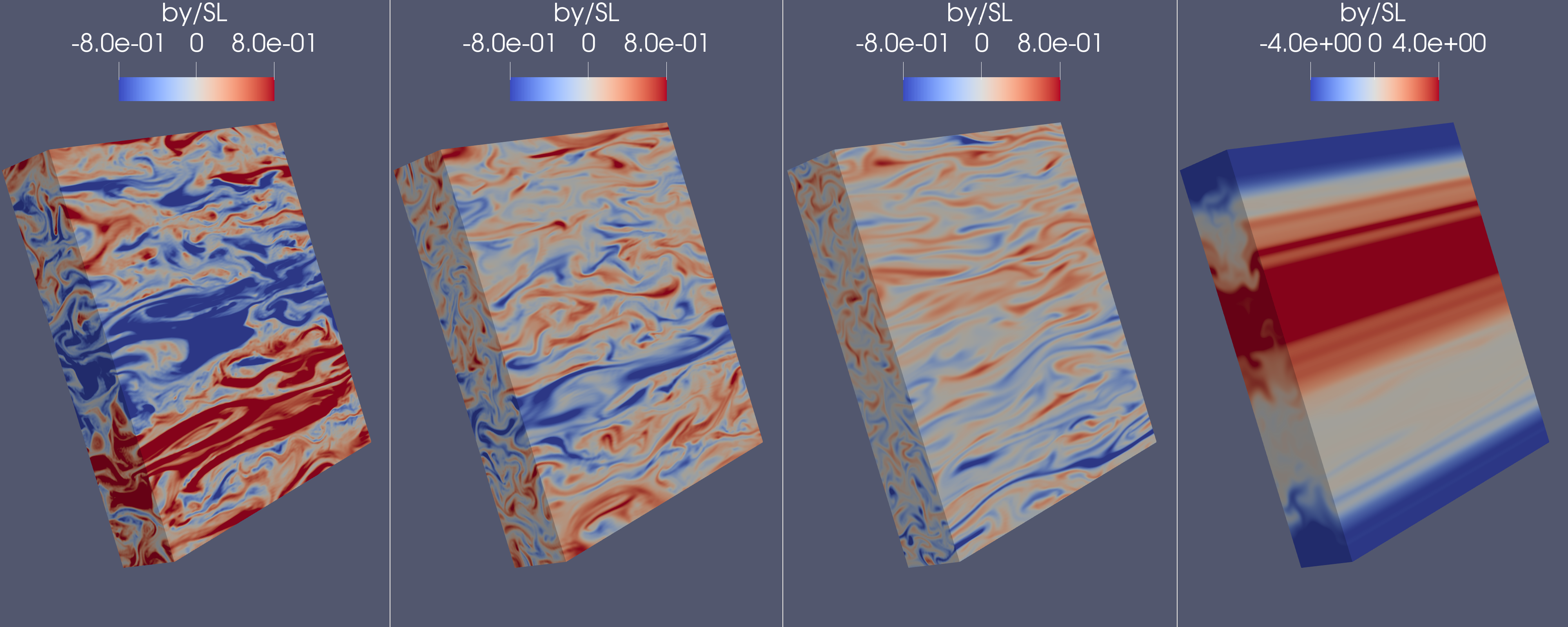} 
   \caption{Same as Fig.\ref{fig_vx_strat} but for the longitudinal magnetic field $b_y$. }
   \label{fig_by_strat}
\end{figure*}

The typical amplitude and lengthscale of each component of the velocity and magnetic fields were studied as a function of $PrN^2/S^2$. Figures \ref{fig_vx_strat} and \ref{fig_by_strat} respectively show chosen snapshots of the radial velocity field $v_x$ and of the longitudinal magnetic field $b_y$ at $PrN^2/S^2=1.2\times10^{-2}, 1.2\times10^{-1}, 4.8\times10^{-1}$ and $2.5$. Inspection of figure \ref{fig_vx_strat} enables to clearly see the reduction
of the amplitude of $v_x$ (note that the color scale has been modified for each snapshot) together with the lengthscale decrease in both the $x$ and $z$ directions, as stratification increases. 

The effect of stable stratification is less drastic on the longitudinal magnetic field, as shown in Figure \ref{fig_by_strat}. In particular, we note that the amplitude of $b_y$ stays approximately constant as $PrN^2/S^2$ increases. The lengthscale decrease in the $x$ and $z$ directions is however still visible. 
In the last case at $PrN^2/S^2=2.5$, turbulence is lost after a few tens of shear timescales and we can see that the axisymmetric component of $b_y$ grows initially and does not seem to be subject to the MRI. This is compatible with criterion \ref{eq_limit} of the linear analysis applied to $\Omega/S=1/Ro=2/3$, which tells us that the instability should vanish for $PrN^2/S^2\geq4/9$. The simulation at $PrN^2/S^2=4.9\times10^{-1}>4/9$ is however found to be still turbulent, at least over the calculated $2000 \, S^{-1}$. This may be explained by the various approximations used in the linear analysis and in particular the comparison of the maximum growth rate to the viscous damping rate to establish Eq.\ref{eq_limit}. For example, as shown in \cite{Guilet15} in the standard MRI case, viscosity may only reduce the growth rate without entirely suppressing the instability. We finally note that both for $v_x$ and $b_y$, the typical wavenumber in the $y$-direction $k_y$ seems quite unaffected by the stable stratification, and is clearly smaller than $k_x$ and $k_z$. This validates the use of the approximation $k_y << (k_x, k_z)$ in the linear analysis.

\begin{figure}[h!]
\centering
   \includegraphics[width=0.5\textwidth]{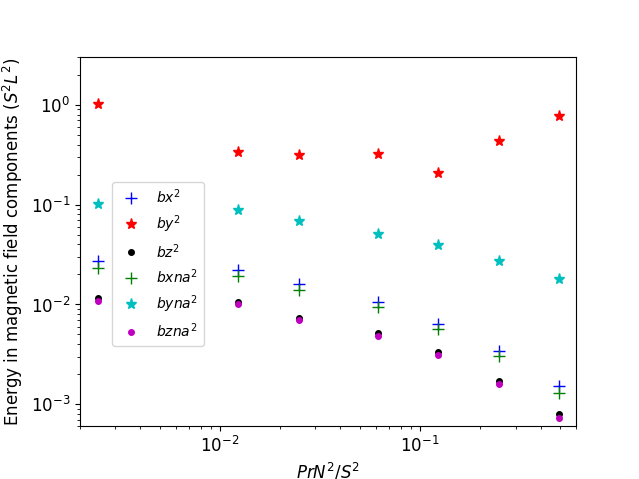} 
   \includegraphics[width=0.5\textwidth]{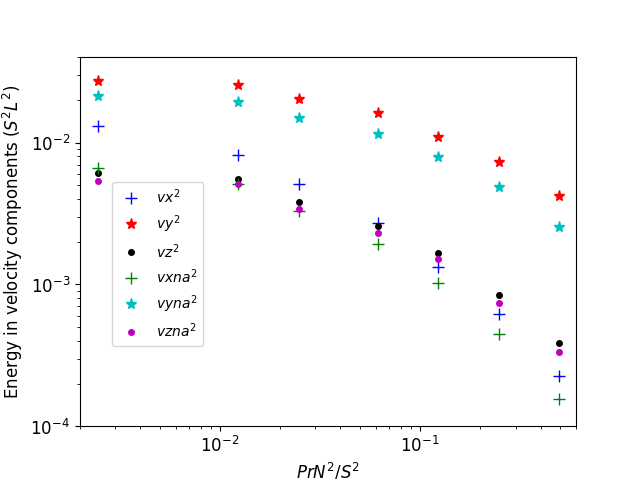}
   \caption{Mean values of the magnetic (top) and kinetic (bottom) energies as a function of stratification measured by $PrN^2/S^2$. Both the total and non-axisymmetric (denoted by \emph{na}) components are shown in each panel. The magnetic energy is always strongly dominant. Note that this plot is made at $Re=1529$ because more simulation points were available but trends are similar at $Re=765$.}
   \label{fig_b2v2}
\end{figure}

 A more quantitative estimate of the reduction of the amplitude of the various components of the flow and field with stratification is shown in Figure \ref{fig_b2v2}. The time-averaged values of each component of the total and non-axisymmetric magnetic and velocity fields are plotted as a function of $PrN^2/S^2$. We recover the strong reduction in the radial velocity $v_x$, while the total $b_y$, dominated by its axisymmetric part, remains mostly unaffected. 
 In general, we find that even at a small level of stable stratification, strong anisotropies are present in the system as the longitudinal $y$-component of both the magnetic and velocity fields dominates the other two components. Anisotropies are amplified by the stable stratification as the vertical velocity and magnetic fields are more strongly affected by an increase in $PrN^2/S^2$ than the horizontal components. 
 We can estimate the scaling of the components of the flow and field which are involved in the calculation of the Maxwell and Reynolds stresses, namely the non-axisymmetric parts of $v_x, v_y, b_x$ and $b_y$. We find that the fluctuations of $v_x$ scale approximately as $(PrN^2/S^2)^{-0.4}$, while the non-axisymmetric $v_y$ scales as $(PrN^2/S^2)^{-0.2}$ (the same scaling is true for the total $v_y$). As $v_y$ is the dominant component of the velocity field (as seen on the figure), we find that $v^{rms}$ follows the same scaling with stratification. Similar scalings are found for the non-axisymmetric parts of $b_x$ and $b_y$, namely $b_x^{\prime} \propto (PrN^2/S^2)^{-0.4}$ and $b_y^{\prime} \propto (PrN^2/S^2)^{-0.2}$.

\begin{figure}[h!]
\centering
   \includegraphics[width=0.5\textwidth]{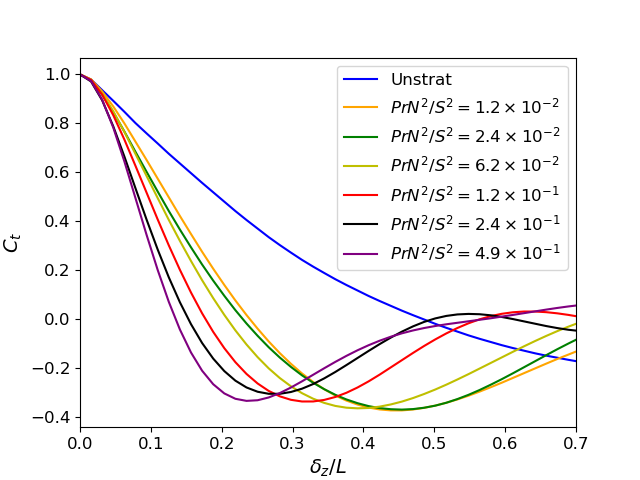}
\includegraphics[width=0.5\textwidth]{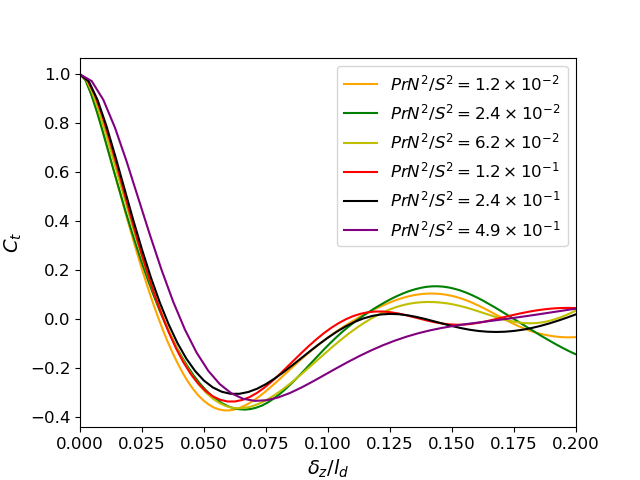}
   \caption{Transverse autocorrelation function $Ct(\delta_z)$ as a function of $\delta_z$, for different values of the stratification measured by $PrN^2/S^2$. Top: with $\delta_z$ on the x-axis, bottom: rescaling the $x$-axis by $l_d=v^{rms}/S$.}
   \label{fig_corvxz}
\end{figure}

Let us now discuss more quantitatively the lengthscale reduction as stratification increases. It is interesting that stratification does not affect only the velocity lengthscale in the $x$-direction but also strongly acts on the horizontal wavenumber $k_z$. To better distinguish the effect on the scales in both directions and give more quantitative estimates, we performed an analysis of the two following autocorrelation functions, referred respectively as the longitudinal and transverse autocorrelation functions in turbulence studies \citep{Batchelor53} and applied to MRI studies for example in \cite{Guan09, Davis10,Walker16}:% Guan et al. 2009; Davis et al. 2010):

\begin{equation}
Cl(\delta_x)=\langle v_x(\oa {r} + \delta_x \, \oa {e_x})\, \,v_x (\oa {r} )\rangle /\langle v_x(\oa {r})^2\rangle 
\label{eq_corvxx}
\end{equation}
and
\begin{equation}
Ct(\delta_z)=\langle v_x(\oa {r} + \delta_z \, \oa {e_z})\, \,v_x (\oa {r} )\rangle /\langle v_x(\oa {r})^2\rangle 
\end{equation}
where $\langle .\rangle $ denotes an average over the whole volume. 
Figure \ref{fig_corvxz} shows the transverse autocorrelation function as a function of $\delta_z$, calculated at one particular time for simulations at different values of $PrN^2/S^2$.

To calculate the mean correlation length from these auto-correlation functions, one would have to integrate those functions from $0$ to $\infty$ (or at least to a value of $\delta_x$ or $\delta_z$ above which the integral does not significantly vary anymore). As we are here mostly interested in the scaling of the typical lengthscales as a function of stratification, we can just attempt to find a rescaling of the $x$-axis so that all curves coincide. In our simulations, several timescales could be relevant: the eddy turnover time $v^{rms}/l$, the modified buoyancy timescale $\kappa/l^2N^2$ and the dynamical timescale $1/S$. By equating the first two timescales, we find a typical lengthscale above which turbulent eddies will feel the effects of stratification $l_s=(v^{rms}\kappa/N^2)^{1/3}$. By equating the turnover and dynamical times, we find a lengthscale for which the flow adjusts to the shear timescale $l_d=v^{rms}/S$. We note that the full turbulent velocity amplitude $v^{rms}$ was adopted here, instead of only the $x$-component $v_x^{rms}$ (which may better represent the turnover timescale), as it was found to produce better results for the scaling. 

On the bottom panel of figure \ref{fig_corvxz}, we rescaled $\delta_z$ by a relevant lengthscale so that all curves would collapse. We find that in the $z$-direction, a rescaling by $l_d$ seems to be the best adjustments to our simulation results. As $v^{rms}$ was found to decrease as $(PrN^2/S^2)^{-0.2}$ and $l_z \propto l_d$, $l_z$ follows the same scaling as  $v^{rms}$. In the $x$-direction, it is a combination of $l_s$ and $l_d$ which is a better indicator of $l_x$ (see figure \ref{fig_corvxx} in appendix \ref{sec_appendix_strat} for illustration). We thus find that $l_x$ is proportional to $\sqrt{l_s l_d}$, which actually gives a slightly stronger dependence of $l_x$ on stratification, closer to a power law index equal to $-0.3$. 
Note that a similar scaling, namely $l_x \propto (PrN^2/S^2)^{-1.6/5}$, is obtained from a time scale balance $l_x/v_y = t_{BM}$ with $t_{BM} = (\kappa/N^2) \times (l_z^2/l_x^4)$
valid as $l_x \ll l_z$ and using the scalings of $v_y \sim v_{rms}$ and $l_z$ found in the simulations. This time balance is inspired by the similar balance
in diffusionless stably stratified turbulence \citep{Lindborg06}. The scaling of $v_x$ then follows as mass conservation, $div(\overrightarrow{v})=0$, leads 
to
$v_y/l_y = v_x/l_x$, with $l_y$ largely unaffected by stratification.

We can conclude from these results that the system adapts to stable stratification by modifying the typical flow lengthscale in the $z$-direction so that the turnover timescale of the turbulent eddies remains close to the shear timescale. We also find that the Alfv\'en frequency, defined as $\omega_A=k_y b_y^{axi}$,
remains also nearly constant
since $k_y$ and the axisymmetric part of $b_y$ stay quite unaffected by the stratification.
These two properties are reminiscent of the linear analysis where
$k_z$ adapts so that the growth rate and the Alfv\'en frequency remain relatively constant as stratification increases (as shown in Figure \ref{linear_a} of appendix \ref{sec_linear}). 

This indicates that, despite the  stable                                           
stratification,  the non-linear 
system adapts to maintain an efficient self-sustained process by reducing the flow lengthscale in the $z$-direction and by maintaining
the amplitude of the axisymmetric longitudinal field.
However, a constant $v^{rms}/l_z$ and a decreasing $l_z$ still lead to a smaller turbulent energy in the system.

The various reductions in amplitude and lengthscales of the flow and field will directly affect the angular momentum and chemical element transport. We now more specifically focus on the effects of stratification on these quantities.

\subsection{On the angular momentum transport}
\label{sec_strat_AM}

\begin{figure}[h!]
\centering
   \includegraphics[width=0.5\textwidth]{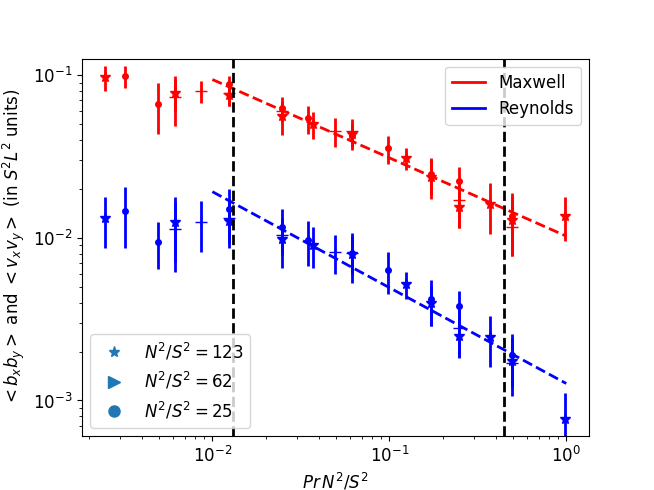}
   \caption{ Absolute value of the Maxwell and Reynolds stresses $\langle b_x b_y\rangle $ and $\langle v_x v_y\rangle $ as a function of $PrN^2/S^2$. Three sets of simulations varying $Pr$ are performed, each set with different value of $N^2/S^2$. The impact of stratification is clearly controlled by the product $Pr N^2/S^2$. The slope for the Maxwell stress is $-0.58$ and for the Reynolds stress $-0.62$, compatible with similar exponents. The leftmost vertical dashed line indicates the value of $PrN^2/S^2$ at which stratification is expected to start to be active and the rightmost dashed line indicates where the linear analysis predicts stability.}
   \label{fig_stress_N}
\end{figure}

We systematically measured the mean Reynolds and Maxwell stresses  for  different stratification levels. The results are plotted on Fig.\ref{fig_stress_N}, as a function of $PrN^2/S^2$. As expected from the main balance in the temperature equation, we verify that all points at various values of the pair ($N/S$, $Pr$) but same value of the combined parameter $PrN^2/S^2$ collapse on the same value for both the Reynolds and Maxwell stresses. We also observe that Maxwell stresses largely dominate the Reynolds stresses in all cases computed here. Independently of the level of stratification, the ratio of Maxwell to Reynolds stresses is always found to be of the order of $9-10$. This was already observed in unstratified MRI dynamo systems \citep[as initially shown by][]{Hawley95} and holds true in our stratified calculations.

In Fig.\ref{fig_stress_N}, we can identify three different regimes: stratification measured by $PrN^2/S^2$ first seems inoperative on the mean stresses, then imposes some reduction and then kills the turbulence. The first regime is obtained for $PrN^2/S^2 < 10^{-2}$. After this value, the stresses decay following a power law in $PrN^2/S^2$ and turbulence is then lost for $PrN^2/S^2\approx1$. This last transition can be related to the linear analysis, which predicts stability for $PrN^2/S^2\geq4/9$. We verify here that turbulence is indeed lost between $PrN^2/S^2=5\times10^{-1}$ and $PrN^2/S^2=1$ (corresponding to the last point of figure \ref{fig_stress_N}) in our calculations, approximately compatible with the predicted value.

We interpret the first two regimes to be distinguished by the ratio of the typical time scale of buoyancy $t_{BM}=\kappa k^2/ N^2$ compared to the dynamical timescale that we take as the inverse of the shear rate $t_S=1/S$. From the linear analysis, we know that the instability may choose a typical lengthscale to accommodate for the effects of stratification. We can then estimate that the motions will remain unaffected by the stable stratification as long as $t_{BM}>t_S$ at all scales contained in the simulation box. This implies that the wavenumber $k$ should be such that $k^2 > N^2/\kappa S$, or in adimensioned variables: 
$$k^2 L^2 > Pr \,\, Re \,\, \frac{N^2}{S^2} $$
If we take $k$ as the smallest wavenumber fitting in the box in the $x$-direction $k=\pi/L_x$, we find that all scales can escape the stabilizing effects of buoyancy thanks to thermal diffusion as long as: 
$$Pr \frac{N^2}{S^2} < \frac{1}{Re} \,\,\frac{\pi^2 L^2}{L_x^2}$$
 For our choice of $L_x=L$, we find a transitional value of $Pr N^2/S^2=1.3\times 10^{-2}$. Again, this limit seems compatible with the value at which a change of slope is observed in Fig. \ref{fig_stress_N}.

In between these two limits, we find that the stresses decrease with the level of stratification. Both stresses approximately follow the same powerlaw, namely $\propto (PrN^2/S^2)^{-0.58}$ for the Maxwell stress and $\propto (PrN^2/S^2)^{-0.62}$ for the Reynolds stress, with small variations of these power law indexes when the Reynolds number is varied (see appendix \ref{sec_param}). These exponents are compatible with the respective variations of the non-axisymmetric parts of $v_x$, $v_y$, $b_x$ and $b_y$ which were established in the previous section, namely $(v^{\prime}_x, b^{\prime}_x) \propto  (PrN^2/S^2)^{-0.4}$ and  $(v^{\prime}_y, b^{\prime}_y) \propto (PrN^2/S^2)^{-0.2}$. This indicates that the correlations are well approximated by the product of the amplitude of each field and that there is consequently no strong effect of stratification on the spatial correlation itself between the field components.

\subsection{On the chemical element transport}

\begin{figure}[h!]
\centering
   \includegraphics[width=0.5\textwidth]{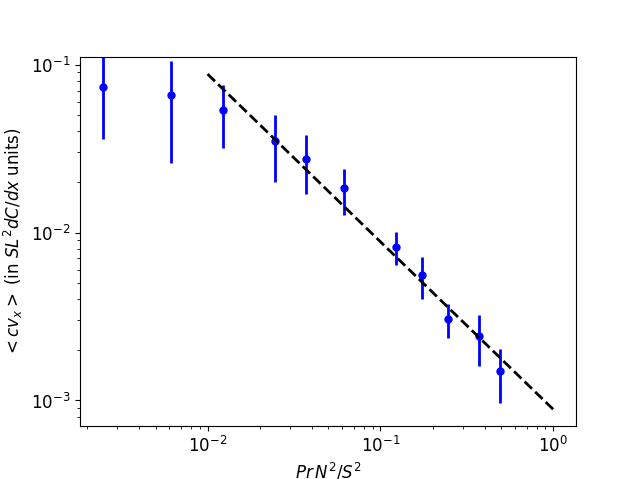}
   \caption{Transport of chemical elements $\langle c v_x\rangle $ as a function of $PrN^2/S^2$, with a fixed value of $N^2/S^2$ and varying $Pr$. The Reynolds number is here fixed to $Re=765$. Slope of the scaling is $-1.05$.}
   \label{fig_cvx_N}
\end{figure}

 We now turn to investigate the behaviour of the concentration flux $\langle c v_x\rangle $, mimicking the turbulent transport in the radial direction of the abundance of a chemical element. As the radial velocity field is strongly affected by the stable stratification, with a typical scaling $v_x \propto (PrN^2/S^2)^{-0.4}$, we also expect a strong decrease in the value of $\langle c v_x\rangle $ as a function of stratification. We find that the effect is even more drastic than expected since $\langle c v_x\rangle  $ is found to be almost exactly inversely proportional to $PrN^2/S^2$. This is shown in Fig.\ref{fig_cvx_N} where the mean value of the flux of chemical elements $\langle cv_x\rangle $ is plotted as a function of $PrN^2/S^2$. The fitted curve has a slope almost exactly equal to $-1$ and we note that this exponent does not vary much with the Reynolds number (see appendix \ref{sec_param}).

The decrease with $PrN^2/S^2$ is faster than the exponent found for $v^{rms}_x$ because the amplitude of $c^{rms}$ also significantly decreases as stratification increases (see appendix \ref{sec_appendix_strat} for illustration). We find that $c^{rms}$ scales as $(PrN^2/S^2)^{-0.55}$, so that $\langle c v_x\rangle $ more or less varies as $v^{rms}_x \times c^{rms}$ with stratification, indicating a constant and strong spatial correlation between the two fields. If we consider that the diffusion of chemicals is well described by a turbulent diffusion process with a diffusion coefficient $D_t=\langle c v_x\rangle /(dC/dx)$ and that this diffusion coefficient can be approximated by $(v_x^{rms})^2 \tau_L$ where $\tau_L$ is the Lagrangian correlation time \citep[then assuming Lagrangian homogeneity of the flow, e.g.][]{Lignieres20}, we may estimate this correlation timescale by calculating $c^{rms}/(v_x^{rms}\,dC/dx )$.  This estimated correlation time is found to slightly decrease with stratification with a scaling exponent equal to $-0.13$ (as illustrated in appendix \ref{sec_appendix_strat}). We thus conclude that the decrease in the transport of chemicals is here due to the combined effect of the decrease in the $x$-component of the velocity and in the correlation timescale.

 Interestingly, we find that the transport of chemical elements is then more strongly affected by the stable stratification than the angular momentum transport, due to the fact that $c$ and $v_x$ are both severely affected by an increase in the value of $PrN^2/S^2$ while only the $x$-component of the 
magnetic field is strongly impacted.

\section{Impact of rotation}
\label{sec_rot}
 We now wish to address the question of the impact of rotation on the MRI and the associated transport efficiency. The simulations discussed so far have been performed with a Rossby number $Ro=1.5$, suitable for the context of accretion disks but less justified for stellar interiors where rotation may be much stronger than the shear. In this section, we vary the value of the Rossby number $Ro=S/\Omega$ from  $0.2$ to $1.8$, varying $\Omega$ while keeping $S$ fixed.

\subsection{On the flow and field characteristics}
\label{subsec_ro_geo}

\begin{figure*}[h!]
\centering
 \includegraphics[width=\textwidth]{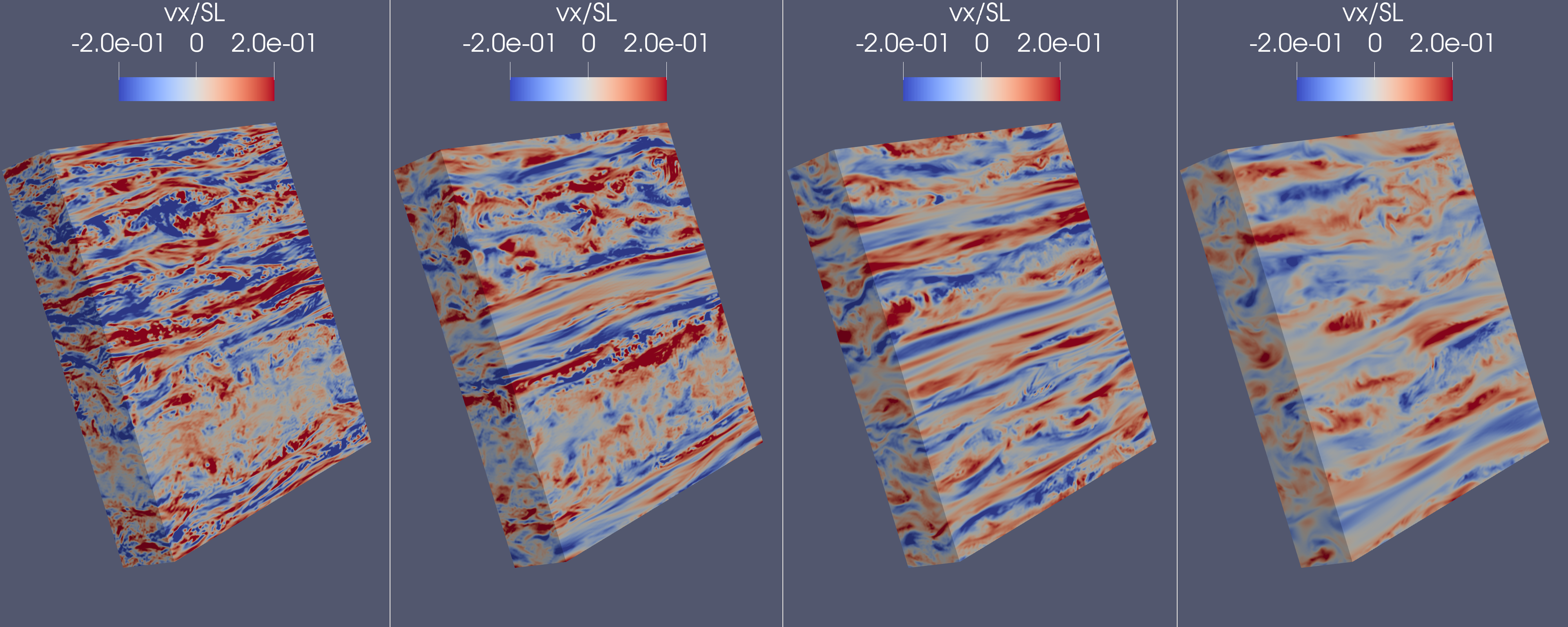}
  \includegraphics[width=\textwidth]{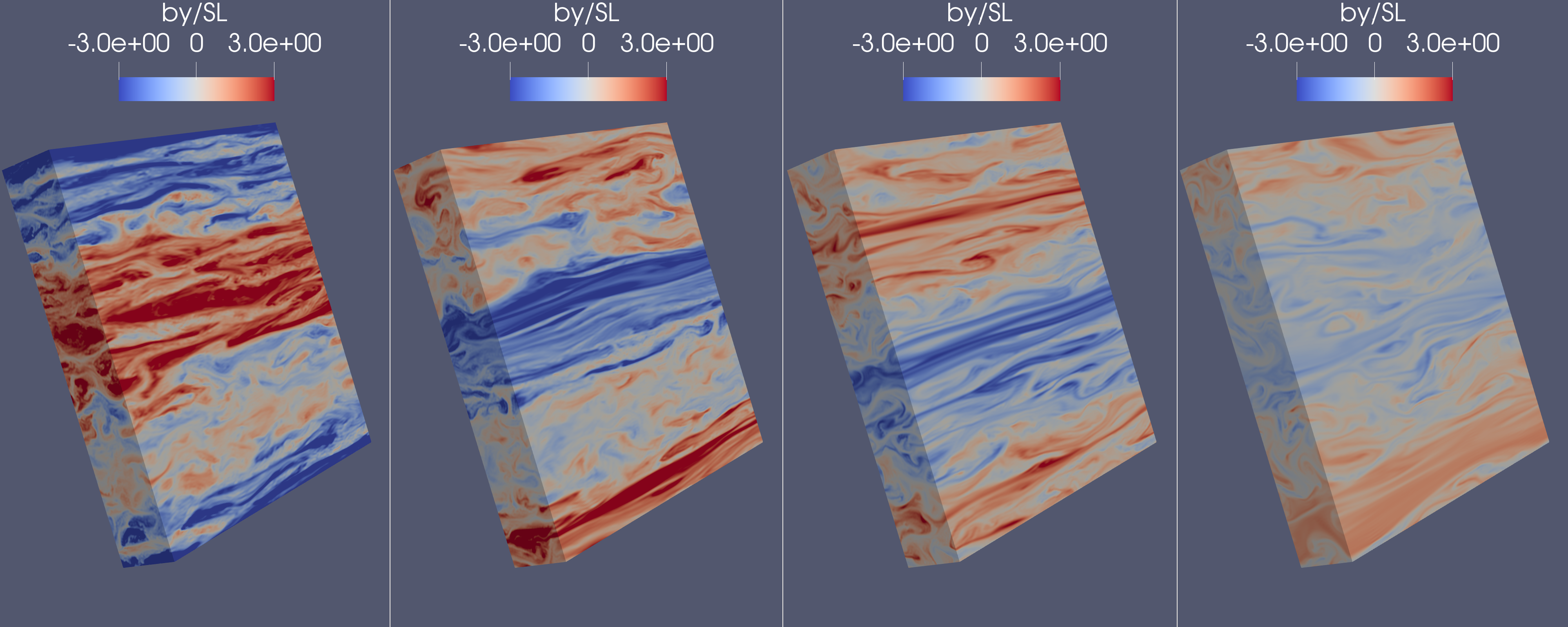}
   \caption{Snapshots of the radial velocity $v_x$ (upper row) and longitudinal magnetic field $b_y$ (lower row) in units of $SL$, for different values of the Rossby number : $Ro=0.2, 0.5, 1 $ and $1.5$ (from left to right). Other parameters are $PrN^2/S^2 = 6\times 10^{-2}, Re = 765$.}
 \label{fig_snapshots5e-4}
\end{figure*}

The first thing to note when studying the effects of the Coriolis force is that they will depend on the amplitude of the stratification. Indeed, the linear analysis of the MRI already showed how tangled were the impacts of rotation and stratification. 
We start by inspecting the case of a small stable stratification: $PrN^2/S^2=6\times 10^{-2}$. On Figure \ref{fig_snapshots5e-4}, we show snapshots of the radial component of the velocity field $v_x$ and of the longitudinal magnetic component $b_y$. We clearly see that rotation mainly changes the lengthscales without significantly modifying the amplitude of the velocity field, while the amplitude of the magnetic field decreases as Ro is increased. Indeed, when $Ro=S/\Omega < 1$ the timescale of the Coriolis force $t_\Omega=1/(2\Omega)$
is smaller than the shear timescale $t_S=1/S$ which was previously
found to constrain the dynamical timescale of the turbulent eddies. 
If we now assume that the eddy turnover time $v^{rms}/l$ adapts to the Coriolis timescale and not to the shear timescale and as $v^{rms}$ does not vary much, this implies that the typical lengthscales should reduce as $Ro$ decreases, as we observe in both the $x$ and $z$ directions.

\begin{figure}[h!]
\centering
   \includegraphics[width=0.47\textwidth]{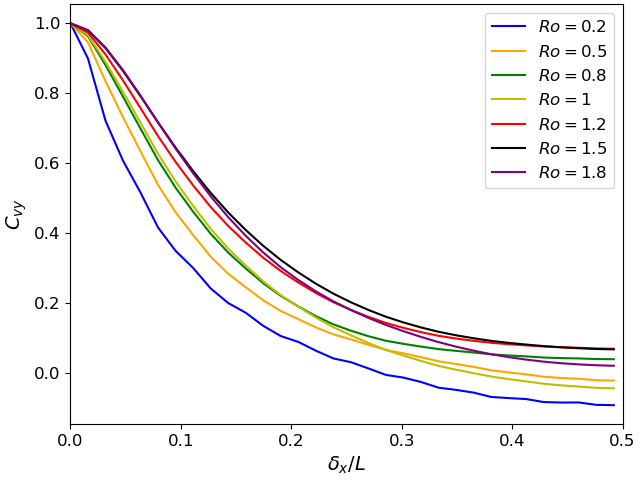}\\
   \bigskip
    \includegraphics[width=0.47\textwidth]{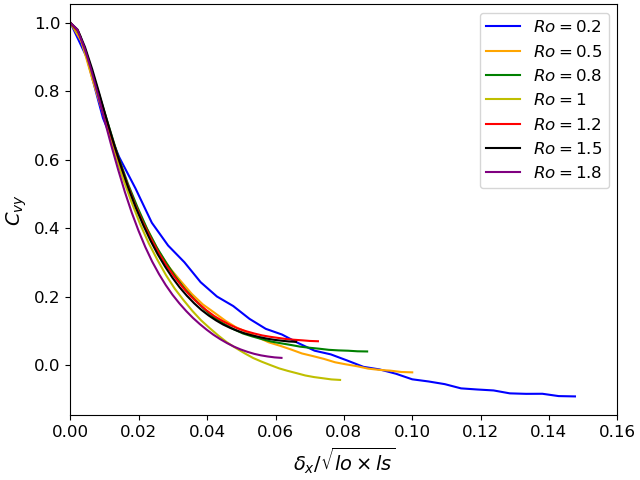}
   \caption{Autocorrelation function of $v_y$ (defined by Eq. \ref{eq_corvxx} but using $v_y$ instead of $v_x$) as a function of $\delta_x$, for different values of the Rossby number, in the weakly stratified case $PrN^2/S^2=6\times 10^{-2}$. Top: as a function of $\delta_x$. Bottom: $\delta_x$ is rescaled by the square root of the product $lo=v^{rms}/2\Omega$ and $ls=(v^{rms} \kappa/N^2)^{1/3}$. As $v^{rms}$ does not vary much with $Ro$, we find that the scale $l_x$ is proportional to $\Omega^{-1/2}$. }
 \label{fig_corel5e-4}
\end{figure}

Applying the same procedure as for the effect of the stratification, we computed the typical scale of the flow by calculating the autocorrelation function of $v_x$ and $v_y$ in the $x$-direction. The results are shown in Fig.\ref{fig_corel5e-4} for the autocorrelation of $v_y$ as a function of $\delta_x$ where it is found that a rescaling of the $x$-axis by $1/\sqrt{\Omega}$ is a good match to our data. We also clearly see on the top panel (where the x-axis is not rescaled) that scales do not vary much between $Ro=1.2$ and $Ro=1.8$, compatible with the fact that rotation is not high enough to
constrain the lengthscales of the flow.

As stratification is increased to $PrN^2/S^2=2.5\times10^{-1}$, we now find that the amplitude of the velocity and magnetic fields is significantly altered when $Ro$ varies, with an increase of both field amplitudes when rotation increases (see Fig. \ref{fig_snapshots2e-3} in appendix \ref{sec_appendix_ro} for illustration). 
This variation in the field and flow amplitudes is understood by the fact that the growth rate of the MRI also increases as $\Omega$ is increased, this being true as soon as $PrN^2/S^2$ is non-zero, as seen in Equation \ref{eq_growth}. 
For small values of the stratification, we thus expect the effects of rotation to act mostly on the typical scale of the motions, thus not significantly affecting the amplitude of the velocity field. However, as shown in the linear analysis, when stratification is ignored, rotation still plays a role on determining the value of the Alfv\'en frequency of the most unstable mode $\omega_A^2=\Omega S(1-S/(4\Omega))$. We can then expect a higher amplitude of the magnetic field when rotation increases, which is consistent with the magnetic field snapshots of Fig.\ref{fig_snapshots5e-4}. This would imply that the Reynolds stresses should be only marginally affected by rotation if stratification effects are small, while the Maxwell stresses would tend to a higher value for smaller $Ro$ in unstratified cases. On the contrary, at a higher level of stratification, rotation injects more energy in the system as the growth rate of the MRI increases with rotation. We then expect both the Maxwell and Reynolds stresses to increase with rotation. We try to confirm these expectations by focusing on the stresses in the next section.

\subsection{On the angular momentum transport}

\begin{figure}[h!]
   \includegraphics[width=0.5\textwidth]{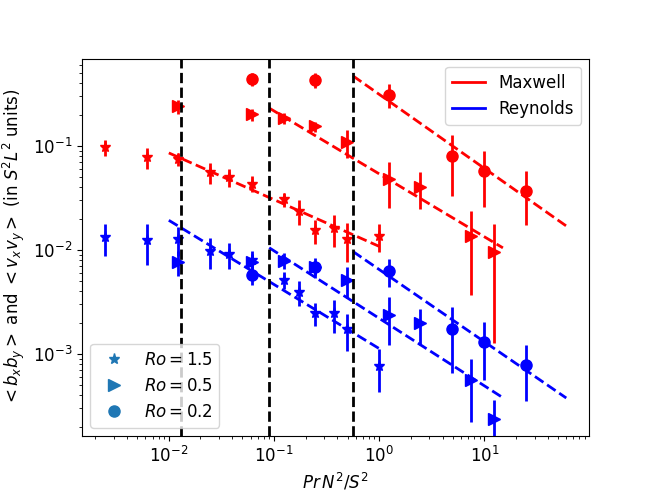}
   \includegraphics[width=0.5\textwidth]{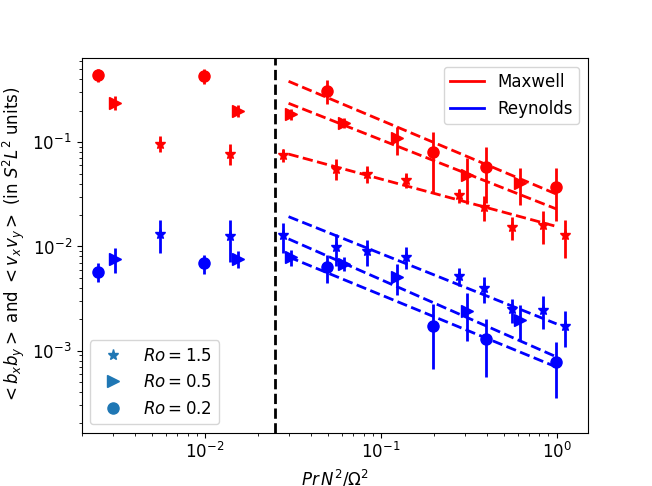}
   \caption{Maxwell and Reynolds stresses for 3 different values of $Ro$, as a function of $PrN^2/S^2$ (top) and $PrN^2/\Omega^2$ (bottom)}.
   \label{fig_stress_3ro}
\end{figure}

We first plot the mean Maxwell and Reynolds stresses as a function of stratification measured by $PrN^2/S^2$, for different values of the Rossby number. We recall that this plot was done for $Ro=1.5$ in Figure \ref{fig_stress_N} and we here simply overplot the same curve for $Ro=0.2$ and $Ro=0.5$. The results are shown on the top panel of Figure \ref{fig_stress_3ro}. 

As expected, we find that the effects of stratification start to be visible on the stresses at higher and higher values of $PrN^2/S^2$ as rotation increases ($Ro$ decreases). We also confirm that, when $N^2$ tends to $0$, the mean Reynolds stresses tend to the same constant value for all $Ro$ while the Maxwell stresses tend to higher values for higher $Ro$. 

Following a similar reasoning as in section \ref{sec_strat_AM}, we can determine the value of $PrN^2/S^2$ above which stratification starts to play a role on the motions. We can do so by comparing the buoyancy timescale to the Coriolis timescale and expect that as long as $t_{BM}>t_{\Omega}$, the system will remain mostly unaffected by the stable stratification. This would now lead to scales such that:
 $$k^2 L^2 >  Pr\frac{N^2}{S^2} \, Re \, \frac{Ro}{2}$$ 
For the $Ro=3/2$ case, the smallest available wavenumber $k$ was considered to be $\pi/L_x$, i.e. corresponding to the box size. We now observe that as $Ro$ decreases and even for small stratification levels, the lengthscale is already reduced by a factor $\sqrt{Ro}$, so that it is now more relevant to choose $k=\pi/(\sqrt{Ro}L_x)$ as the smallest available wavenumber in the box when stratification does not act.
Transition is thus estimated to occur at:
$$Pr\frac{N^2}{S^2}> \frac{2}{ReRo} \, \frac{\pi^2 L^2}{ Ro \, L_x^2}$$
If we take an approximated transition at $10^{-2}$ for $Ro=1.5$, we get at transition at $9\times10^{-2}$ for $Ro=0.5$ and at $5.6\times10^{-1}$ for $Ro=0.2$. Vertical dashed lines indicating these values for the three Rossby numbers are plotted on the top panel of Fig.\ref{fig_stress_3ro} and indeed seem to coincide fairly well with the change of slope of the various curves. The transition can also be written as:
 $$Ro^2\, Pr\frac{N^2}{S^2}=Pr\frac{N^2}{\Omega^2}>\frac{2}{Re} \, \frac{\pi^2 L^2}{L_x^2}$$
This shows that the transition is actually controlled by $PrN^2/\Omega^2$. The bottom panel of Figure \ref{fig_stress_3ro} where the $x$-axis is now $PrN^2/\Omega^2$ illustrates this, the transition estimate being simply around $PrN^2/\Omega^2=2.5\times 10^{-2}$. Moreover, after this transition, we find that the scaling of the stresses with the stratification gives a similar power index for the three different values of $Ro$.

We now compare various simulations at different $Ro$ at the same value of the stratification measured by $PrN^2/\Omega^2$ to determine the effect of rotation on the stresses. Choosing a value of $PrN^2/\Omega^2=6\times 10^{-2}$, i.e. in the domain where stratification acts, we find a scaling in $Ro^{-0.86}$ for the Maxwell stress and $Ro^{0.22}$ for the Reynolds stress. The plot showing the scaling for different values of the stratification and Reynolds numbers is shown in appendix \ref{sec_param}.

When combined with the previous scaling with $PrN^2/S^2$ (which is the same as the scaling with $PrN^2/\Omega^2$ in the previous section where $Ro=S/\Omega$ is fixed), we thus get the following relationships for the Maxwell and Reynolds stresses:

\begin{eqnarray}
\frac{\langle b_xb_y\rangle }{L^2S^2} &\propto& (PrN^2/\Omega^2)^{-0.58}(S/\Omega)^{-0.86}\\
&=&(PrN^2)^{-0.58}S^{-0.86}\Omega^{2.02}
\end{eqnarray}
\begin{eqnarray}
\frac{\langle v_xv_y\rangle }{L^2S^2} &\propto& (PrN^2/\Omega^2)^{-0.62}(S/\Omega)^{0.22}\\
&=&(PrN^2)^{-0.62}S^{0.22}\Omega^{1.02}
\end{eqnarray}

\noindent so that the effect of rotation is always to increase the stresses, the Maxwell stress being more strongly enhanced than the Reynolds stress. This is consistent with rotation increasing the growth rate of the MRI and the most unstable Alfv\'en frequency.

We also note in the scaling that the effect of the shear is different on the Reynolds and Maxwell stresses. 
To relate our results to previous MRI unstratified calculations where the impact of the shear parameter was investigated, we can study the ratio of the Maxwell to Reynolds stresses, as done for example by \cite{Masada12} for a large number of different simulations. This ratio was then compared to estimates determined using the linear properties of the MRI. In particular, \cite{Pessah06} established that even in the non-linear regime, this ratio is well approximated by $\langle b_x b_y\rangle/\langle v_x v_y\rangle = (4-Ro)/Ro$.
Their study was conducted for the standard MRI, with an imposed vertical magnetic field. \cite{Masada12} then showed that the scaling with $Ro$ seemed to remain valid for any configuration of the MRI, including zero-net flux cases similar to the ones performed in this work. 
They however showed that $\langle b_x b_y\rangle/\langle v_x v_y\rangle  \propto (4-Ro)/Ro$, with a proportionality coefficient which can be quite different from $1$. In our calculations, we find $\langle b_x b_y\rangle/\langle v_x v_y\rangle  \approx 3.5 \times (4-Ro)/Ro$ to be a good approximation of the stress ratio for almost all cases. Quite remarkably, we show in addition to the previous studies that the scaling is valid for almost all levels of stratification (see appendix \ref{sec_appendix_ro} for illustration).

We conclude from this section that the transport due to the MRI is here favored by a strong rotation, partly since the instability grows faster with rotation, and that the dominance of the Maxwell stresses in the transport is even stronger at large rotation rates. 
\subsection{On the chemical element transport}

\begin{figure}[h!]
\centering
  \includegraphics[width=0.5\textwidth]{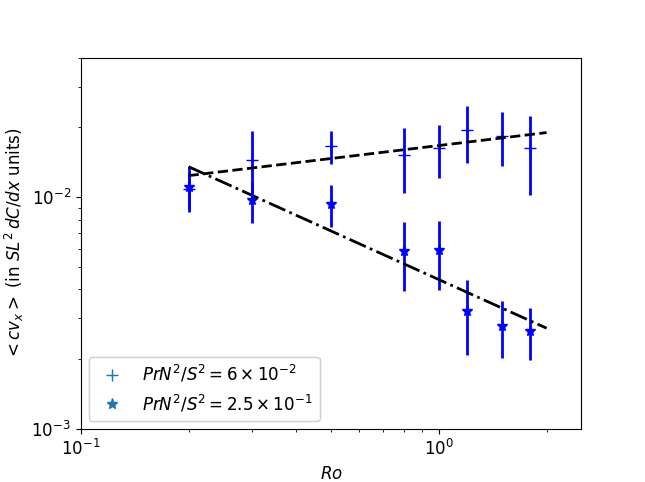}
   \caption{Transport of chemical elements $\langle cv_x\rangle $ as a function of the Rossby number (where $S$ is fixed while $\Omega$ is varied), for two values of the stratification measured by $PrN^2/S^2$. The dashed and dashed-dotted lines represent fitted powerlaw scalings in both cases. Note that these scalings are obtained with fixed $PrN^2/S^2$ and are therefore different than that of equation \ref{eq_scaling_cvx} obtained with fixed $PrN^2/\Omega^2$ in the highly stratified regime.} 
   \label{fig_cvx_ro}
\end{figure}

As far as the chemical element transport is concerned, it is quite interesting to compare the scaling in Rossby first for two fixed values of $PrN^2/S^2$, emphasizing the difference between a weakly and strongly stratified case. Figure \ref{fig_cvx_ro} shows the results for two values of $PrN^2/S^2=6\times 10^{-2}$ and $2.5\times10^{-1}$, i.e. the same values as the ones chosen to discuss the effects of rotation on the flow and field in section \ref{subsec_ro_geo}. 

In this figure, it is quite clear that the effect of rotation depends on the level of stratification. In the weakly stratified case, the transport of chemicals is almost unaffected by rotation while it is strongly enhanced by rotation in the more stably stratified case. We note that if rotation is sufficiently strong (i.e. at small Rossby numbers), the transport of chemicals is similar between the two levels of stratification. This is compatible with the discussion above where we highlighted the fact that strong rotation builds spatial scales small enough to escape the stabilizing effect of stratification. If again we consider that the chemical transport is well approximated by $\langle c v_x\rangle  \approx (v_x^{rms})^2 \, \tau_L \, dC/dx$, we can estimate the Lagrangian correlation time $\tau_L$ at both levels of stratification. In both cases, we find that the Lagrangian correlation time decreases with rotation (illustrated in figure \ref{fig_cvrms_ro} of appendix \ref{sec_appendix_ro}). This is expected as smaller and smaller lengthscales are built when rotation increases. As a consequence, at small stratification, the slight decrease of $v_x^{rms}$ with $Ro$ is compensated by the increase in $\tau_L$, leading to an almost constant $\langle cv_x\rangle $. In the strongly stratified case, $\langle cv_x\rangle  \approx(v_x^{rms})^2 \tau_L$ decreases with $Ro$ as the decrease in $v_x^{rms}$ overcompensates the increase of $\tau_L$.

When stratification is further increased, the scaling with Rossby does not significantly vary anymore. By now choosing a value of $PrN^2/\Omega^2$ in the regime where stratification is effective at all Rossby numbers, we find a scaling such that $\langle c v_x\rangle \propto Ro^{0.92}$. A plot showing the scaling for different values of the stratification and Reynolds numbers is shown in appendix \ref{sec_param}.
Combined with the previous scaling in $PrN^2/\Omega^2$, we then get, for $c$ and $v_x$:

\begin{eqnarray}
\frac{\langle cv_x\rangle }{SL^2\,dC/dx} &\propto& (PrN^2/\Omega^2)^{-1}(S/\Omega)^{0.92} \nonumber \\ 
&=&(PrN^2)^{-1}S^{0.92}\Omega^{1.08}
\label{eq_scaling_cvx}
\end{eqnarray}
so that both the shear and rotation increase the value of the chemical element transport, at sufficiently high stratification, with an almost linear relationship between the transport and $\Omega$.

Finally, we see here that again rotation acts differently on the transport of chemicals and angular momentum, with the transport of AM being more enhanced by rotation than $\langle cv_x\rangle $. Again, this is understood by the fact that rotation increases both the growth rate and the Alfv\'en frequency of the most unstable mode, both effect playing a role at increasing the magnetic field amplitude and thus the Maxwell stresses. On the contrary, only the enhanced growth rate at high stratification helps rotation to increase the transport of chemicals.

\section{Impact of the Reynolds and Prandtl numbers}
\label{sec_reynolds}

From the two previous sections, in the range of values of $PrN^2/\Omega^2$ where both stratification and rotation impact the angular momentum and chemical transport, the following relationships are found, first for the Maxwell stress:
\begin{equation}
\frac{\langle b_x b_y\rangle }{L^2 S^2} =\alpha(Re, Pm)\, \,  Ro^{-0.86} \, (Pr N^2/\Omega^2)^{-0.58}
\label{eq_max1}
\end{equation}

\noindent and for the chemical transport:
\begin{equation} 
\frac{\langle c \, v_x\rangle }{S L^2 \, dC/dx} =\beta(Re, Pm) \, \,  Ro^{0.92} \, (Pr N^2/\Omega^2)^{-1}
\label{eq_cvx1}
\end{equation}

We here recall that $\oab$ is a reduced magnetic field which has the unit of a velocity, so that the factor $4\pi\rho$ does not appear in Eq. \ref{eq_max1}.

We are left with studying the dependencies of coefficients $\alpha$ and $\beta$ as a function of $Re={S}L^2/\nu$ and $Pm=\nu/\eta$. To do so, we calculated additional cases at Reynolds numbers $Re=191, 382, 765, 1529, 2294, 3058$ and $4050$ and at $Pm=2,4$ and $8$. We focus here on a level of stratification $PrN^2/\Omega^2 \in [5\times 10^{-2}, 1]$, i.e. in the range where the scalings on $\alpha$ and $\beta$ previously established hold.

\subsection{Dependency on the Reynolds number}

\begin{figure}[h!]
\centering
   \includegraphics[width=0.5\textwidth]{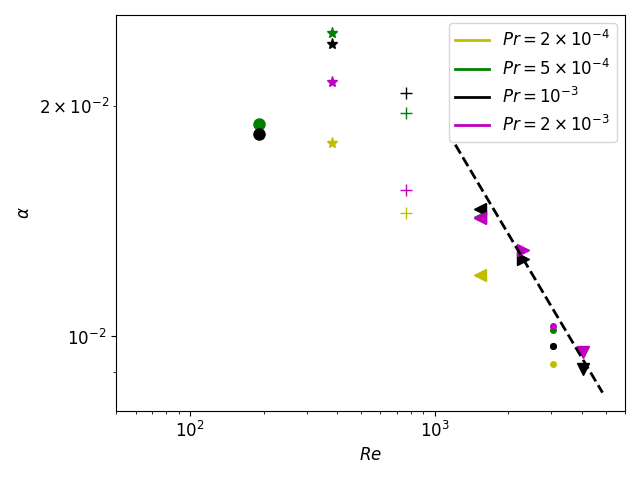}      
    \includegraphics[width=0.5\textwidth]{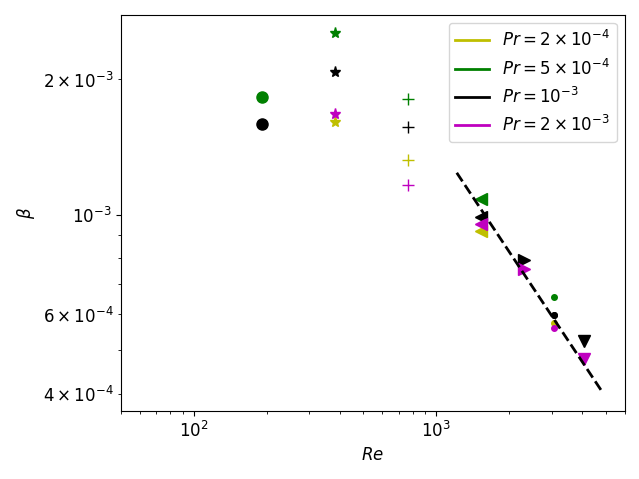}%
   \caption{Transport of angular momentum denoted by $\alpha$ in equation \ref{eq_max1} (top) and chemical element transport denoted by $\beta$ in equation \ref{eq_cvx1} (bottom) as a function of $Re$, for 4 values of $Pr$ and at fixed $N/S=11$ and $S/\Omega=1.5$ (i.e. in the range of $PrN^2/\Omega^2$ where the previous scalings were established) and $Pm=16$. Fits shown by dotted lines are based on the last three values of $Re$ and for $Pr\geq 5\times10^{-4}$. The exponent is $-0.58$ for $\alpha$ and $-0.81$ for $\beta$.}

   \label{fig_re}
\end{figure}

On figure \ref{fig_re}, we plot the behaviour of coefficients $\alpha$ and $\beta$ as a function of the Reynolds number, for 4 values of the stratification (varying $Pr$, keeping $N/S=11$), at $S/\Omega=1.5$ and at $Pm=16$. In both cases, we have clues that some asymptotic regime is reached at high values of the Reynolds number. 
The fact that the points at $Pr=10^{-3}$ and $Pr=2\times10^{-3}$ collapse for $Re>10^3$ indicate that the fit established in $PrN^2/\Omega^2$ for both the Maxwell stress and the transport of chemicals remains valid at $Re>10^3$, even though it was established at $Re=765$. 

On this figure, we see that the data points seem to collapse only at large values of $Re$ and $PrN^2/\Omega^2$. This can be understood through the following argument, which was already suggested by \cite{Prat16} and \cite{Garaud17} in the context of transport of chemicals by shear instabilities: when the Reynolds number is too small and stratification is too low, the typical scale of turbulent eddies is comparable to the size of the simulation box. In our cases, we find a similar behaviour: for a fixed value of $PrN^2/\Omega^2=2.78\times10^{-1}$, we need to have $Re>\approx 2\times 10^3$ to reach typical lengthscales well below the size of the box in the $x$-direction. Above this value, turbulence adapts its typical lengthscale to accommodate for the stratification and we then do not expect the box size in the $x$-direction to affect the results anymore (see appendix \ref{sec_appendix_re}). In this asymptotic regime and as indicated by the black dotted lines, a power law can be fitted through our simulation points for $Re>2\times 10^3$. The power law for the angular momentum transport $\alpha$ is $Re^{-0.58}$, and $Re^{-0.81}$ for the chemical element transport denoted by $\beta$.

When stratification is ignored, \cite{Fromang10}, using MRI simulations in shearing boxes similar to the ones performed in this work, showed that at fixed $Pm$, the value of the angular momentum transport done by the Maxwell stress (also denoted $\alpha$ in the context of accretion disks) remained roughly independent of the Reynolds number. However, when stratification is important, we find different results and values of $\alpha$ and $\beta$ reveal to be strongly impacted by $Re$. As we shall see in the next section, our scalings with $Re$ are compatible with transport coefficients which are asymptotically independent of the viscosity, which is quite crucial for possible extrapolations to stellar regimes.

\subsection{Dependency on the magnetic Prandtl number}

\begin{figure*}
   \includegraphics[width=0.5\textwidth]{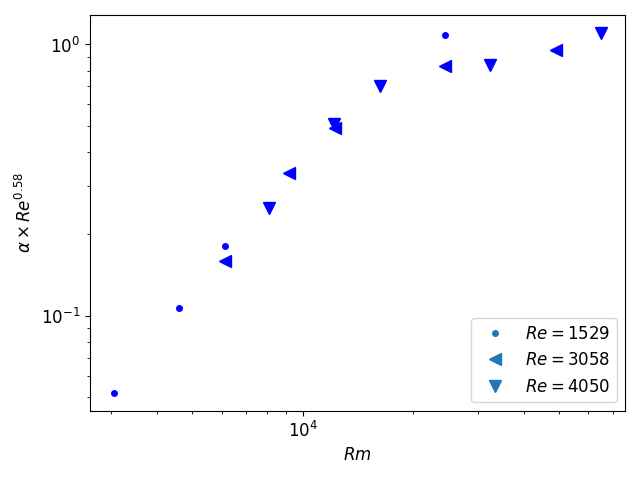}       
   \includegraphics[width=0.5\textwidth]{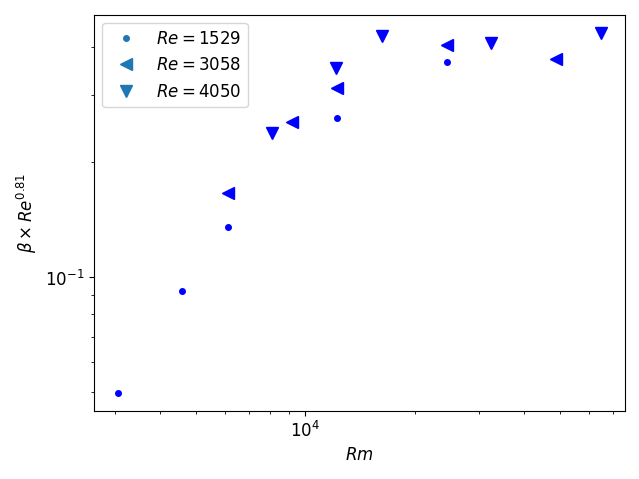}
 \caption{$Re^{0.58} \times \alpha$ and $Re^{0.81} \times \beta$ as a function of $Rm=RePm$, for a stratification level $PrN^2/\Omega^2=2.78\times10^{-1}$. An asymptotic regime seems to be reached for high values of $Rm$, for which there is no dependency on $Rm$ anymore.}
   \label{fig_pm}
\end{figure*}

We end this study with the dependency of the $\alpha$ and $\beta$ coefficients on the magnetic Prandtl number $Pm$. We note that all cases calculated fail at producing a dynamo when the value of $Pm=1$ is used. This is a known feature of MRI-driven dynamo calculations that sustaining turbulence at $Pm$ close to $1$ is a challenge. We calculated cases for $Re=1529, 3058$ and $4050$ at $PrN^2/\Omega^2=2.78\times10^{-1}$ (i.e. in the asymptotic regime discussed above) at $Pm=2, 4$ and $8$. For the case at the lowest $Re$ and $Pm$, turbulence is also lost after a few hundreds of shear timescales. For the other cases, we find sustained turbulence, at least for the $2000$ shear timescales calculated.

Since the dependency of $\alpha$ and $\beta$ with $Re$ were established in the previous section, we now focus on the dependency of $\alpha \times Re^{0.58}$ and $\beta \times Re^{0.81}$ on $Pm$. The results are shown in figure \ref{fig_pm} for the angular momentum transport (left) and the chemicals (right), which are plotted as a function of $Rm=RePm$.

We find a clear trend towards an asymptotic regime as both $Re$ and $Pm$ are increased.
When plotted against $Rm=Pm Re$, both curves reach an asymptotic regime around $Rm=2\times 10^4$, where $\alpha$ and $\beta$ tend to values of order 1. In stellar interiors, the value of $Pm$ is typically less than $1$, so that an asymptotic regime reached at high $Pm$ is not particularly relevant for stellar application. However, if the asymptotic regime is reached in terms of $Rm$ instead of $Pm$ then our results may also hold for small $Pm$ values. Dynamo calculations being out of reach for the moment in regimes at small $Pm$ and high $Rm$, we can only speculate that our asymptotic regime is still valid for stellar interiors. We will come back to applications of this work in the stellar context in the next section.

A similar trend to an asymptotic regime was also seen in the unstratified calculations of \cite{Shi16} where large aspect ratio boxes were used and an asymptotic regime was also reached for the transport of angular momentum in $Pm$. An asymptotic regime was also reached in \cite{Guilet22,Held22} for shearing boxes less elongated in the vertical direction, in which the MRI-driven dynamo does not show the same kind of oscillating mean axisymmetric azimuthal field as described in Section~\ref{sec:dynamo}. Their trend to the asymptotic regime was nonetheless different compared to our results: the transport was approximately independent of $Re$ and the asymptotic regime applicable only above a relatively high value of $Pm$.

\section{Discussion}
\label{sec_discu}

From the previous sections, we now have a complete estimate of the dependencies of the transport of angular momentum and chemical elements by the stratified MRI on all the relevant parameters likely to play a role in stellar interiors. We place ourselves in a regime where $3\times10^{-2}<PrN^2/\Omega^2<1$, $S/\Omega<2$, $Re>2\times10^3$ and $Rm>2\times10^4$, where the scalings established before hold. We can first wonder if this regime is applicable to stellar interiors. An estimate of the value of  $PrN^2/\Omega^2$ was previously discussed in \cite{Gouhier21} and it was established that regimes where $PrN^2/\Omega^2<1$ could easily be reached in pre-main sequence stars or outside the degenerate cores of subgiant stars, owing to the very small value of the Prandtl number in those regions. This is considering the Brunt-V\"ais\"al\"a frequency due to thermal gradients and ignoring the effects of $\mu$-gradients, which may be quite important in red giant stars for example. The opposite limit $PrN^2/\Omega^2> 3\times10^{-2}$ is not always realized in stars, again because of small $Pr$ values. However we noted that this limit where all the vertical scales are significantly smaller than $\sqrt{\kappa/N}$ is controlled by  the  box size in the $x$-direction of our numerical set-up: if larger boxes were considered, the transport scaling would be extended to lower $PrN^2/\Omega^2$ values. In stars, MRI-unstable regions will be confined on the extent of the radial gradient of $\Omega$. If, as expected, this length scale is much larger than $\sqrt{\kappa/N}$, this unstratified limit will not be relevant.

We here focus on the regime where the above scalings were established, namely $3\times10^{-2}<PrN^2/\Omega^2<1$.

In the regime, we can then rewrite equation \ref{eq_max1} as:

\begin{equation}
\frac{\langle b_x b_y\rangle }{L^2S^2} =\alpha\, \,  Ro^{-0.86} (\Omega^2/N^2)^{0.58}(\kappa/\nu)^{0.58} \, (\nu/L^2 S)^{0.58}
\label{eq_max2}
\end{equation}

\noindent with $\alpha \approx 1$.

As stated above, this already shows that the dependency in the viscosity $\nu$ vanishes in this expression, which is very encouraging for possible applications to stellar interiors whose typical Reynolds numbers are out or reach numerically. 

We can also then express the stress as an equivalent turbulent viscosity by stating that 
\begin{equation}
\nu_t=\frac{\langle b_x b_y\rangle }{dV_y/dx}=\frac{\langle b_x b_y\rangle }{S}
\end{equation}
We then get:

\begin{eqnarray}
\nu_t&=&\alpha\, \,  Ro^{-0.86} (\Omega^2\kappa/N^2)^{0.58} \,  (L^2 S)^{0.42}\nonumber \\
&=&\alpha \,\, S^{-0.44} \Omega^{2.02} \, (\kappa/N^2)^{0.58} \, L^{0.84}
\label{eq_nu}
\end{eqnarray}

We can compare the scaling of $\nu_t$ as a function of stratification with previous studies aiming at determining the efficiency of angular momentum transport by the MRI. First phenomenological arguments were given in \cite{Spruit99} where the turbulent resistivity was assumed to adjust such that the instability condition given by the linear analysis \ref{res3} (when $Pm$ smaller than $1$) will be just satisfied (i.e. the system places itself just above the stability limit). This gives a turbulent viscosity proportional to $\kappa/N^2$, corresponding to a steeper dependence with stratification compared to equation \ref{eq_nu}. Recent simulations of the azimuthal MRI have also been performed by \cite{Meduri24} in global simulations. In the last study, a variation of $\nu_t$ with $(\kappa /N^2)^{0.25}$ is found, again different from what is deduced from our current study. These simulations were global and a mean radial transport was computed from their calculations. It is well possible that the transport efficiency outside the equatorial region may be affected by the stable stratification differently than at the equator, producing a different scaling with $\kappa/N^2$.

Let us now compare our expression of $\nu_t$ with that established for the transport by Tayler-unstable fields \citep{Fuller19}: 
\begin{equation}
\nu_{TI}=\alpha^3 r^2 \Omega^3 \, N_{eff}^{-2}
\end{equation}
\noindent with $\alpha \approx 1$ to fit the observations of the rotation rate of post-main sequence stars and $N_{eff}^2=\frac{\eta}{\kappa} N_T^2 + N_\mu^2$. If we consider only the thermal gradients here (as we do in our calculations), the dependency of $\nu_{TI}$ to the stratification reads $\nu_{TI} \propto \kappa/N^2$. We see that the MRI transport is then less affected by the stable stratification, as well as the minimum shear required for instability, which is proportional to $N^2/\kappa$ for the MRI and  $(N^2/\kappa)^{1.25}$ for the TI prescribed by \cite{Fuller19}. Finally, we note that we still have a dependency on the shear rate $S$ which is absent in the TI prescription of \cite{Fuller19}. This may be interpreted by the fact that the MRI is a shear-driven instability while the Tayler instability is current driven and the shear is only here to produce the necessary azimuthal field for instability. 

We note that if we slightly simplify the aforementioned functional dependency (namely if we assume $0.86\approx 1$ and $0.58\approx 0.5$), we get the following simplified expression for the Maxwell stress:
\begin{equation}
\langle b_x b_y\rangle  =\alpha\, \, \Omega^{2} \, S^{0.5} \, L \, \, (\kappa/N^2)^{0.5}
\end{equation}

\noindent which can also be rewritten in terms of a turbulent viscosity:
\begin{eqnarray}
\label{eq_nu2}
\nu_t&=&\alpha \,\, \Omega^2 \, S^{-0.5} \, L \,\, (\kappa/N^2)^{0.5}\\ \nonumber
&=&\alpha \,\, (\Omega/S)^2 \,\, (\kappa S^2/N^2)^{0.5} \,\, (SL^2)^{0.5}
\end{eqnarray}
We here recover a dependency on $\kappa Ri^{-1}$ 
involving the Richardson number $Ri=N^2/S^2$ 
which was also thought to be the relevant parameters to characterize the hydrodynamical transport due to shear instabilities \citep{Zahn92, Prat13, Garaud17}. This emphasizes the similarities with the transport by shear instabilities though keeping in mind that rotation here enhances the transport by the Maxwell stress, contrary to shear instabilities for which strong shears compared to the rotation rates are needed \citep[see][where GSF is shown to be dominant over shear instabilities when rotation dominates]{Chang21}.

\begin{figure}[h!]
\centering
   \includegraphics[width=0.5\textwidth]{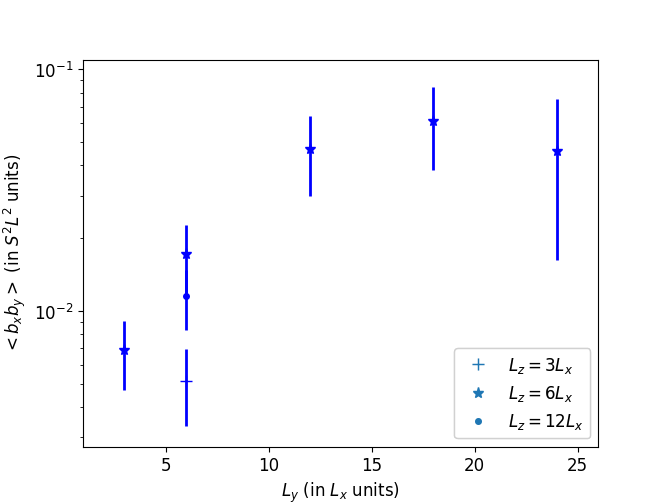}
   \caption{Maxwell stress as a function of the domain size in longitude $L_y$. The 3 points at $L_y/L_x=6$ correspond to different values of $L_z$. $L_x$ was kept fixed and equal to $L$ in these calculations.}
   \label{fig_ly}
\end{figure}

In expression \ref{eq_nu2}, we see that our turbulent viscosity (or our Maxwell stress) seems to be dependent on some lengthscale $L$ of the problem. This was already observed in unstratified MRI simulations \citep[e.g.][]{Fromang07, Simon12} where a dependency in $L^2$ was reported. However, as the aspect ratio of the box was increased in \cite{Simon12}, a convergence of the transport was reached. In our calculations, we do observe that magnetic structures at the size of the simulation box in $y$ and $z$ are always present, even at high $Re$ (see for example the snapshots of $b_y$ in appendix \ref{sec_appendix_re}). This was in fact true for all cases calculated, as seen in the snapshots of Fig. \ref{fig_by_strat} and \ref{fig_snapshots5e-4}. This reveals the important role of the axisymmetric component of $b_y$ in the dynamo loop, as all cases indeed showed a $y$-component of the magnetic field strongly dominated by its $m=0$ contribution. 

In a similar way as the studies performed by \cite{Simon12}, we tried to explore the dependency of the Maxwell stress on the aspect ratio, by varying $L_y$ and $L_z$ independently, while keeping $L_x$ unchanged. Indeed, we do not expect the box size in the direction of gravity to play a significant role on our results when stratification is strong enough, since small scales governed by the level of stratification will fix the scale in this direction, much smaller than $L_x$. This is indeed what we find when we vary $L_x$, keeping $L_y$ and $L_z$ fixed at sufficiently high Reynolds number and high stratification (see appendix \ref{sec_appendix_re} for illustration). When $L_y$ is increased, we do find an indication of convergence of the Maxwell stress while the convergence when varying $L_z$ is less clear. This is shown in figure \ref{fig_ly} where we changed the box size in $L_y$ from $L_y=3 L_x$ to $L_y=24 L_x$ and where for $L_y=6 L_x$, we varied $L_z$ from $3 L_x$ to $12 L_x$.

It is quite encouraging that the stress may become independent of $L_y$ as this is the longitudinal extent of the box. When applied to stars, there is no reason for any physical process to determine a spatial scale significantly smaller than the stellar radius in this direction. However, in the $z$-direction, we may be limited by our local box simulation which may be a good approximation of the equatorial dynamics where the directions of gravity and rotation are perpendicular but becomes less relevant when a significant component of the rotation vector parallel to gravity starts to appear. Physical processes associated with this component may act to limit
the magnetic stresses, thus introducing a finite vertical scale in the stress expression.

If we now discuss the transport of chemical elements, we can rewrite in the same way equation \ref{eq_cvx1} as:

\begin{equation}
\frac{\langle cv_x\rangle }{SL^2\, dC/dx}=\beta \, \, Ro^{0.92}\,(\Omega/N)^2 (\kappa/\nu) \, (\nu/ S L^2)^{0.81}
    \label{eq_beta}
\end{equation}

\noindent with $\beta \approx 0.4$.
Again, this expression is not incompatible with a simplified dependency:
\begin{eqnarray}
\frac{\langle cv_x\rangle }{SL^2 \, dC/dx}&=&\beta \, \, Ro\,(\Omega/N)^2 (\kappa/\nu) \, (\nu/ S L^2)\nonumber \\
&=&\beta\, \Omega \, \,\kappa/(L^2N^2)
    \label{eq_beta2}
\end{eqnarray}

\noindent which can be rewritten in terms of a turbulent dissipation $D_t=\langle cv_x\rangle /\left(dC/dx\right)$ as:
\begin{equation}
D_t=\beta\, \, S \,\Omega \,\frac{\kappa}{N^2} 
    \label{eq_beta3}
\end{equation}

We note that this is exactly the same expression that \citet{Spruit99} established but for the transport of angular momentum (i.e. for the turbulent viscosity, quite different in our case than $D_t$).
Equation \ref{eq_beta3} remarkably suggests that the diffusion coefficient $D_t$ is again independent of the viscosity and this time also of the scale of the problem (by contrast with the AM transport).
This feature makes it particularly interesting for possible extrapolations to stellar values. We argue that stratification is responsible for setting a lengthscale for the velocity field, much smaller than the simulation box, and increasing the Reynolds number will not strongly modify this typical scale. This can be seen on the snapshots of $v_x$ in Fig. \ref{fig_refields} of appendix \ref{sec_appendix_re}. As chemical elements are transported in the vertical direction by $v_x$, it is not surprising that we find scale-independence for the transport $\langle c v_x\rangle $.

Finally, equation \ref{eq_beta3} can be rewritten:
\begin{equation}
\frac{D_t}{\kappa Ri^{-1}}=\beta \, \frac{\Omega}{S}
\end{equation}
so that $D_t/\kappa Ri^{-1}$ is constant for a fixed Rossby number in the asymptotic regime of high Reynolds number and high stratification. This again is reminiscent of what is found for shear instabilities in the high thermal diffusivity regime \citep{Zahn92, Prat13, Prat14}.

\section{Conclusions}
\label{sec_conclu}

In this work, the transport of angular momentum and chemical elements by the stratified MRI dynamo has been studied. A detailed parametric study has been conducted to focus on the effects of the dominant processes present in stellar radiative zones, namely stable stratification measured by a combination of the Brunt-V\"ais\"al\"a frequency and the thermal diffusivity $\kappa/N^2$, rotation measured by $\Omega$ and shear measured by $S$. Further investigations of the effects of the Reynolds and magnetic Prandtl numbers were performed, to finally establish scaling laws between the AM and chemical transport and the relevant parameters. Our setup using the shearing box framework is designed
to study a radial shear in a region close to the equator, so that the shear is in the direction of gravity and the rotation vector is perpendicular to this direction.

In all cases presented in this work, turbulence develops through dynamo action relying on the presence of a magneto-rotational instability of a mean longitudinal magnetic field self-consistently built up in our simulations. We note that the simulations were initialized with a random zero-net flux magnetic field and that our results consequently hold for any differentially rotating weakly magnetized stellar radiative zone. No specific initial condition or magnetic field configuration (above a minimum intensity of the magnetic field) is needed for turbulence to be triggered and transport to be effective. Furthermore, hydrodynamical instabilities are not excited in our setup so that turbulence here necessitates magnetic fields to develop. In those aspects, this work constitutes a promising avenue for the problem of angular momentum and chemical transport in stellar radiative zones.

In unstratified cases at moderate rotation, previous results established in the context of accretion disks are recovered, such as a cyclic variation of the mean magnetic field components, a Maxwell stress always dominating the transport of AM and a ratio of Maxwell to Reynolds stresses depending on the Rossby number $S/\Omega$. When stratification is increased in a regime of high thermal diffusivity relevant to stellar interiors, the typical lengthscales in the $x$ and $z$ directions decrease, along with the amplitude of the velocity field and the non-axisymmetric components of the magnetic field. The mean longitudinal field is however only mildly affected by the stratification, indicating that the self-sustaining process is able to adapt to the stratification. At fixed shear, rotation also significantly alters the flow and field in the stably stratified regime, by decreasing the typical lengthscales and increasing the amplitude of both fields. This is understood by noting that the growth rate of the MRI and the most unstable Alfv\'en frequency both increase with rotation when stratification is high enough. As more energy is injected in the system, both the transport of AM and chemicals increase with rotation, which makes the MRI dynamo a particularly interesting process for fastly rotating radiative interiors, and quite different from previous estimates of the turbulent AM transport related to hydrodynamical instabilities. Finally, we show that our transport coefficients are independent of the viscosity, a very appealing feature for possible extrapolations to stellar regimes for which the typical Reynolds numbers are out or reach numerically.

The established scaling laws for the transport of AM and chemicals in terms of turbulent coefficients as a function $S$, $\Omega$ and $\kappa/N^2$ are deduced from a large number of 3D simulations and could now be incorporated in 1D stellar evolution models to test the efficiency of the MRI dynamo to reconcile asteroseismic constraints with stellar evolution models. Particularly interesting objects for such an application are subgiant and giant stars for which an efficient extraction of angular momentum from the magnetized core is needed to match the observations. Compared to previous estimates of magnetized AM transport based on the Tayler instability \citep[e.g.][]{Spruit02, Fuller19}, the transport by MRI seems less affected by the stable stratification and does not suffer from the same stabilizing effect due to strong rotation. We also establish that the transport of chemicals is always less efficient than that of the AM and that their scaling with respect to the parameters are different. This is interesting for applications to solar-type stars where indeed distinct transport coefficients are needed in stellar evolution models to reproduce both the internal rotation profile and the surface Lithium abundance \citep[e.g.][]{Dumont21,Eggenberger22}.

We note that our study does not include the effect of the compositional gradients which may strongly enhance the Brunt-V\"ais\"al\"a frequency for example in the vicinity of the hydrogen-burning shell in red giant stars. If these gradients were taken into account, the stabilizing effect could not be mitigated by the thermal diffusivity and the system may be stabilized. However, as discussed in \cite{Skoutnev25b}, if a strong magnetic field coming from previous phases of stellar evolution was still existing in this region, magnetic tension could rapidly impose rigid rotation there and the barrier imposed by $\mu$-gradients would not be relevant for the transport.

Our numerical setup is adapted to studying the transport by the stratified MRI dynamo in the equatorial region of a stellar radiative zone. Possible modifications to the established scaling laws could occur if other latitudes were considered at which the rotation vector would have a non-zero component along the shear and gravity directions. In particular, the latitudinal extent of our simulation box, entering the scaling of our Maxwell stress, could well be related to the typical latitude at which this non-zero component of the rotation along the gravity could play a significant role on the transport.
Another important extension of this work would be to focus on a horizontal shear which was shown to be less subject to the stabilizing effects of stratification \citep{Jouve20, Gouhier22}. Numerical simulations aiming at tackling these questions are underway.

\acknowledgements{This work was granted access to the HPC resources of IDRIS and CINES under allocations 2025-A0170413776 and 2026-A0190413776 made by GENCI. The authors also thank Geoffroy Lesur for access to the most recent version of the SNOOPY code and for very fruitful discussions.}

\bibliographystyle{aa.bst}
\bibliography{mybib}

\begin{appendix}

\section{Balance in the temperature equation}
\label{appendix_temp}

\begin{figure*}[h!]
\centering
   \includegraphics[width=\textwidth]{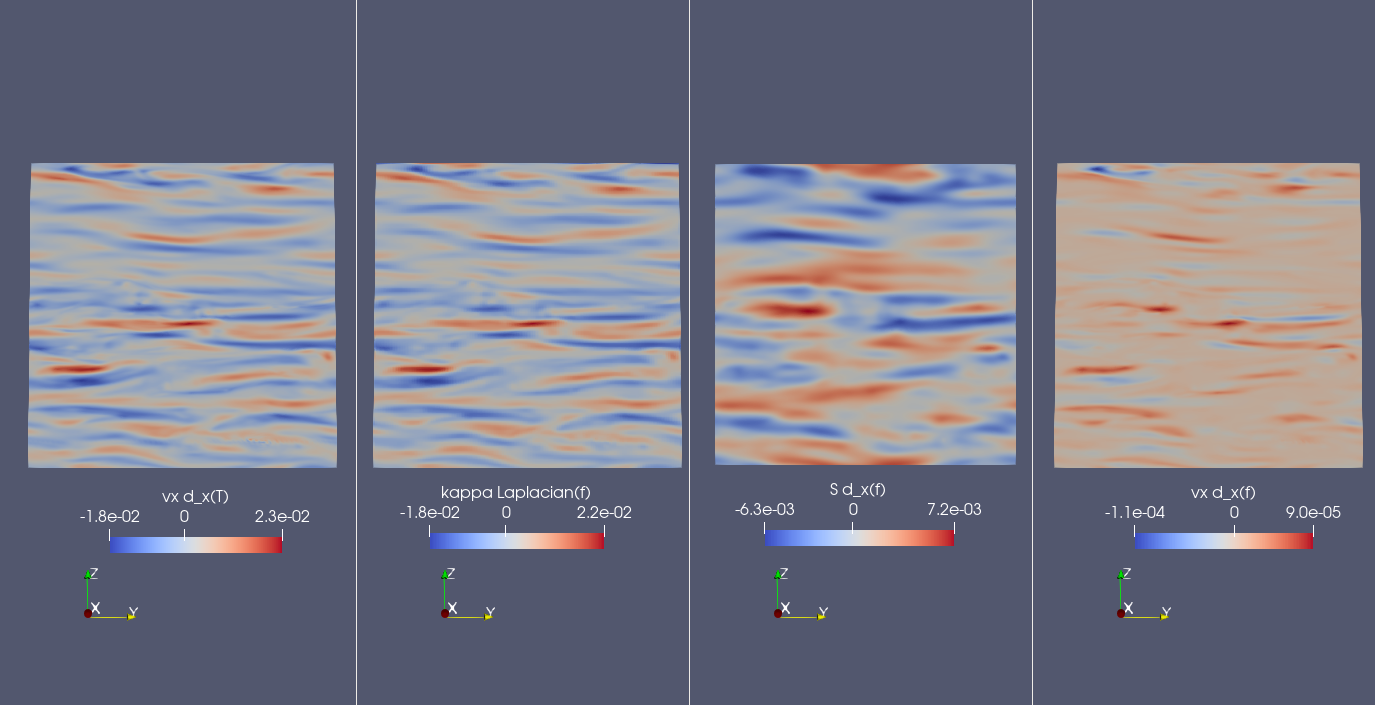}
   \caption{2D slices (mid-plane in x) at a particular time in the simulation of the different terms in the temperature equation in the case at $Pr=4\times 10^{-3}$, ${N/S}=11$. The almost perfect equality between diffusion of fluctuations and advection of mean gradient shows that stratification is controlled by the product of $N^2/S^2$ and $Pr$ only. This holds for all times and all stratified cases.}
   \label{fig_tempeq}
\end{figure*}

We here illustrate the fact that the main balance in the temperature equation \ref{eq_temp} is between the advection against the background temperature gradient $N^2v_x$ and the thermal diffusion of the temperature fluctuations $\kappa \Delta f$. Figure \ref{fig_tempeq} shows $yz$ cuts at mid-plane in $x$ of the various terms in Equation \ref{eq_temp}, at a particular time for a particular stratified case at $Pr=4\times 10^{-3}$ and ${N/S}=11$. The balance is clearly between the first two terms and this holds true for all cases at all times in the simulations. This allows to measure the level of stratification by the parameter $PrN^2/S^2$ instead of $Pr$ and $N/S$ separately.

\section{Linear analysis}
\label{sec_linear}
From the linearized MHD inviscid equations and assuming $X = \hat{X} \exp{\left[\sigma t + i \vec{k}(t)\cdot \vec{x}\right]}$ and $k_y \ll k_x, k_z$, we obtain the following equation for the growth rate $\sigma$:
 \begin{equation}
\label{bousq}
(\sigma^2 +\omega_A^2)^2 + (\sigma^2 +\omega_A^2) \left[2\Omega(2\Omega-S)\frac{k_z^2}{k^2} + \frac{\sigma \omega_g^2}{\sigma + k^2 \kappa} \right] - 4 \Omega^2 \omega_A^2 \frac{k_z^2}{k^2} =0
\end{equation}
\noindent where $\omega_A^2 = \frac{\left(\vec{k}\cdot\vec{B}_0\right)^2}{4\pi \rho_0}$  and $\omega_g^2 = \frac{k_y^2 + k_z^2}{k^2}N^2$.

In the absence of thermal diffusion $\kappa=0$, the previous equation can be solved using the variable $\sigma^2 + \omega_A^2$. We find that $2 \Omega S > N^2$ is the condition for instability. In typical radiative zone conditions where $N \gg \Omega$, this prevents a non-diffusive MRI for Rayleigh stable radial differential rotation $S/\Omega <2$.

By decreasing the amplitude of the buoyancy force, thermal diffusion enables the MRI. In the low-Péclet-number limit $\sigma \ll k^2 \kappa$, the above equation reduces to:
 \begin{equation}
\label{petitpe}
(\sigma^2 +\omega_A^2)^2 + (\sigma^2 +\omega_A^2) \left[2\Omega(2\Omega-S)\frac{k_z^2}{k^2} + \frac{k_z^2}{k^4} \frac{N^2}{ \kappa} \sigma \right] - 4 \Omega^2 \omega_A^2 \frac{k_z^2}{k^2} =0
\end{equation}

Eq.~(\ref{petitpe}) can be viewed as an equation for the two variables $\sigma$ and $\omega_A^2$. Introducing ${\omega}_B^2=\omega_A^2+\sigma^2$, we obtain a related equation:
 \begin{equation}
\label{ms}
 \sigma^2 + \sigma \frac{\omega_B^2}{4 \Omega^2} \frac{N^2}{\kappa k^2} + \frac{\omega_B^2}{4 \Omega^2} \left(\omega_B^2\frac{k^2}{k_z^2} -2 \Omega S \right)  =0
\end{equation}
which can be solved analytically for $\sigma$. A solution $(\sigma,\omega_B^2)$ of Eq.~(\ref{ms}) provides a solution $(\sigma,\omega_A^2=\omega_B^2-\sigma^2)$ of Eq.~(\ref{petitpe}). In addition, if the solution of Eq.~(\ref{ms}) verifies $d\sigma/d\omega_B^2=0$, then the related solution $(\sigma,\omega_A^2=\omega_B^2-\sigma^2)$ of Eq.~(\ref{petitpe}) verifies $d\sigma/d\omega_A^2=0$. Thus, from the fastest growing solution of Eq.~(\ref{petitpe}),
 \begin{equation}
\label{sig_1}
 \sigma =\frac{S}{2} \frac{k_z}{k} \frac{1}{\frac{N^2}{\kappa k^2} \frac{1}{4 \Omega}\frac{k_z}{k}  + 1}
\end{equation}
\noindent reached at 
 \begin{equation}
\label{omA_1}
 \omega_B^2 = \Omega S \frac{k_z^2}{k^2} \frac{1}{\frac{N^2}{\kappa k^2} \frac{1}{4 \Omega}\frac{k_z}{k}  + 1}
\end{equation}
we can deduce the maximum growth rate of the MRI in the low-péclet-number approximation:
 \begin{equation}
\label{sig_2}
 \sigma =\frac{S}{2} \frac{k_z}{k} \frac{1}{\frac{N^2}{\kappa k^2} \frac{1}{4 \Omega}\frac{k_z}{k}  + 1}
\end{equation}
\noindent for 
 \begin{equation}
\label{omA_2}
 \omega_A^2 = \Omega S  \frac{k_z^2}{k^2} \frac{1}{\frac{N^2}{\kappa S k_z^2} \frac{S}{4 \Omega} + 1} \left(1 -\frac{S}{4\Omega}\frac{1}{\frac{N^2}{\kappa S k_z^2} \frac{S}{4 \Omega} + 1} \right)
\end{equation}

The growth rate still depends on the wavenumber components $(k_x,k_z)$. In the limit of small vertical wavelength, the maximum growth rate of the unstratified case, $\sigma_{max} = S/2$ and the corresponding $\omega_A^2 = \Omega S (1 -\frac{S}{4\Omega})$, are recovered.  For a fixed $k_z$, we find that the maximum of $\sigma$ is reached for $k_x=0$ if $k_z^2 \ge \frac{N^2}{2 \Omega \kappa }$, and for $k_x^2 = \left(\frac{N^2}{2 \Omega \kappa }k_z\right)^{2/3} - k_z^2$ for large vertical wavelengths $k_z^2 \le \frac{N^2}{2 \Omega \kappa }$. That is: 
 if  $k_z^2 \ge \frac{N^2}{2 \Omega \kappa }$:
 
 \begin{equation}
\label{res0}
 \sigma =\frac{S}{2} \frac{1}{\frac{N^2}{\kappa k_z^2} \frac{1}{4 \Omega}  + 1}
  \;\; \mbox{and} \;\;
  \omega_A^2 = \Omega S \frac{1}{\frac{N^2}{\kappa S k_z^2} \frac{S}{4 \Omega} + 1} \left(1 -\frac{S}{4\Omega}\frac{1}{\frac{N^2}{\kappa S k_z^2} \frac{S}{4 \Omega} + 1} \right)
\end{equation}
whereas if $k_z^2 \le  \frac{N^2}{2 \Omega \kappa }$:
 \begin{equation}
\label{res1}
 \sigma =\frac{S}{3} \frac{1}{\left(\frac{N^2}{\kappa k_z^2} \frac{1}{2 \Omega} \right)^{1/3}}
 \;\; \mbox{and} \;\;
 \omega_A^2 = \frac{2}{3}\Omega S \frac{1}{\left(\frac{N^2}{\kappa k_z^2} \frac{1}{2 \Omega} \right)^{2/3}} \left( 1- \frac{S}{6 \Omega} \right)
\end{equation}

\begin{figure*}[h!]
\centering
   \includegraphics[width=\textwidth]{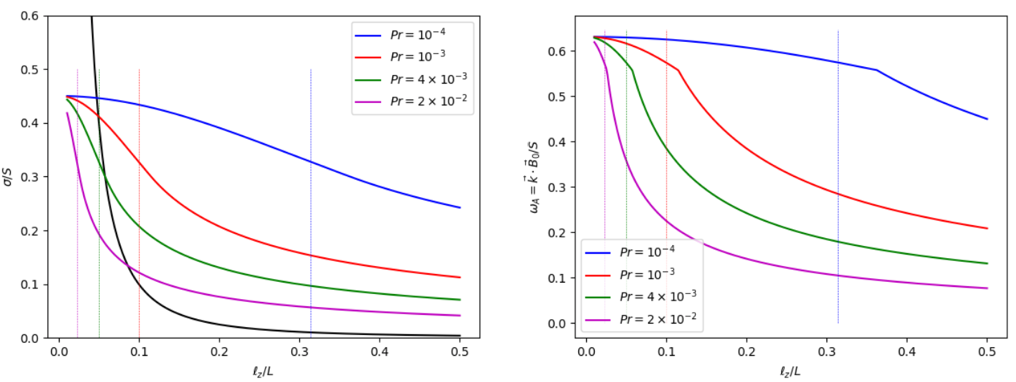}  
   \caption{Maximum growth rate (left) and corresponding Alfv\'en frequency (right) of the MRI in the small-Péclet number regime as a function of the vertical lengthscale of the perturbation for different values of the Prandtl number. The viscous damping rate is also plotted in black. The other parameters are, the Reynolds  $Re= 8\times 10^2$, the Brunt-V\"ais\"ail\"a frequency $N/{S}=11$ and the Rossby number $Ro=S/\Omega = 3/2$. Vertical dotted line mark the vertical length scales that equal the buoyancy lengthscale.}
   \label{linear_a}
\end{figure*}

\begin{figure*}[h!]
\centering
   \includegraphics[width=\textwidth]{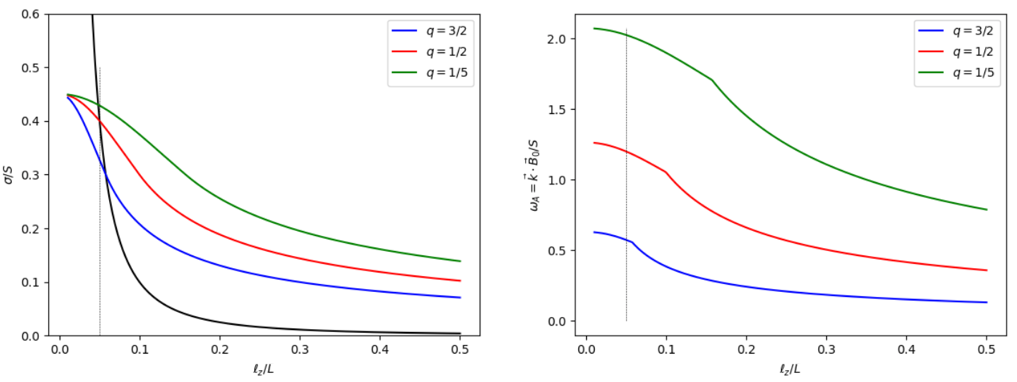}  
   \caption{Maximum growth rate (left) and corresponding Alfv\'en frequency (right) of the MRI in the small-Péclet number regime as a function of the vertical lengthscale of the perturbation for different values of the Rossby number (here denoted by $q$). The viscous damping rate is also plotted in black. The other parameters are, the Reynolds  $Re= 8\times 10^2$, the Brunt-V\"ais\"ail\"a frequency $N/{S}=11$ and the Prandtl number $Pr = 4\times 10^{-3}$.  The vertical dotted line marks the vertical length scale that equals the buoyancy lengthscale.}
   \label{linear_b}
\end{figure*}

As long as the vertical wavelength is small with respect to the buoyancy scale $\ell_B^2 = \frac{\kappa S}{N^2}$, or the Rossby number is small, the growth rate and the Alfv\'en frequency remain close to their unstratified values. But for larger vertical wavelength, both quantities decrease under the effect of the stable stratification. This behaviour is illustrated on Figs.~(\ref{linear_a}) and (\ref{linear_b}) for typical parameters of the numerical simulations. For these parameters, we also find very small differences between these solutions in the low-Péclet-number limit and the maximum growth rates computed using the full dispersion relation Eq.~(\ref{bousq}). 

Viscosity and magnetic diffusivity are expected to stabilize the flow. We can estimate their effects by comparing the maximum growth rate found in their absence with the damping rates $\nu k^2$ or $\eta k^2$. By doing so, we find that the condition for instability is:
 \begin{align}
\label{res3}
\frac{S}{\Omega} \ge \frac{3}{2}\frac{N^2}{\Omega^2} \frac{\nu}{\kappa} \;\; \mbox{if} \;\; P_m \ge 1  \\
\frac{S}{\Omega} \ge \frac{3}{2}\frac{N^2}{\Omega^2} \frac{\eta}{\kappa} \;\; \mbox{if} \;\; P_m \le 1
\end{align}
\noindent this last condition being similar but not equivalent to the one given by \cite{Acheson78} (see also equation (29) of \citet{Spruit99}).

\section{Variation with stratification}
\label{sec_appendix_strat}

\begin{figure*}[h!]
\centering
   \includegraphics[width=0.48\textwidth]{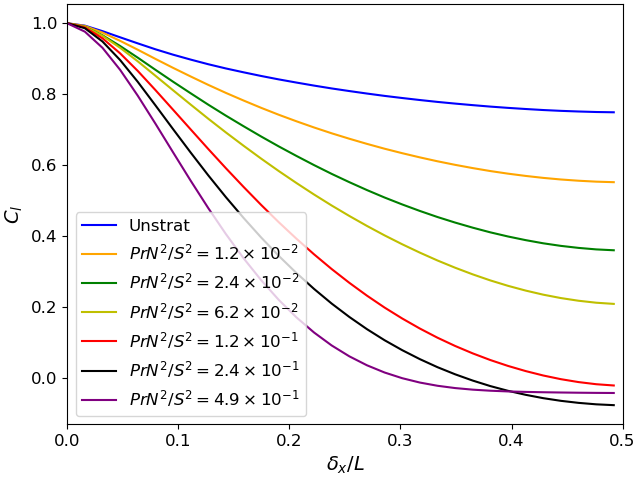}
\includegraphics[width=0.48\textwidth]{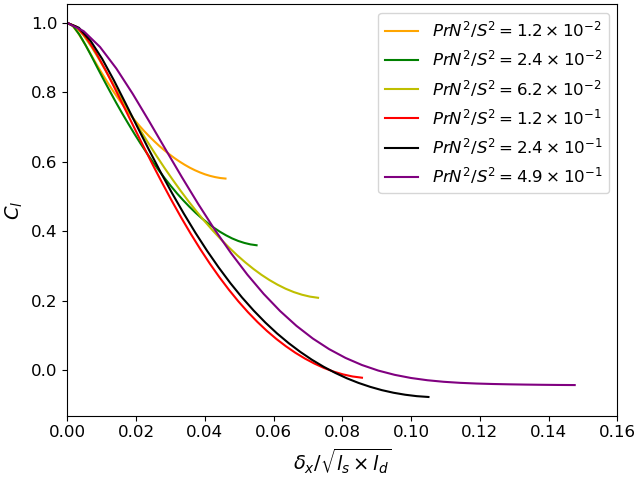}
  \caption{Longitudinal autocorrelation function $Cl (\delta_x)$ as a function of the lengthscale $\delta_x$. Left: with $\delta_x$ on the x-axis, right: rescaling by $(l_d l_s)^{1/2}=(v^{rms}/S\times(v^{rms}\kappa/N^2)^{1/3})^{1/2}$.}
   \label{fig_corvxx}
\end{figure*}

\begin{figure*}[h!]
\centering
   \includegraphics[width=0.49\textwidth]{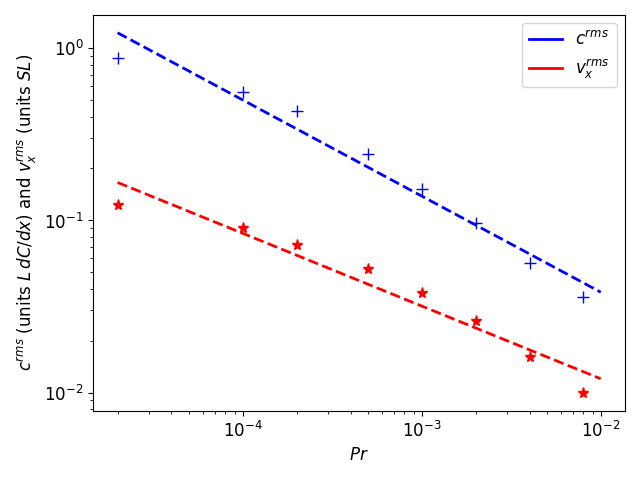} 
  \includegraphics[width=0.49\textwidth]{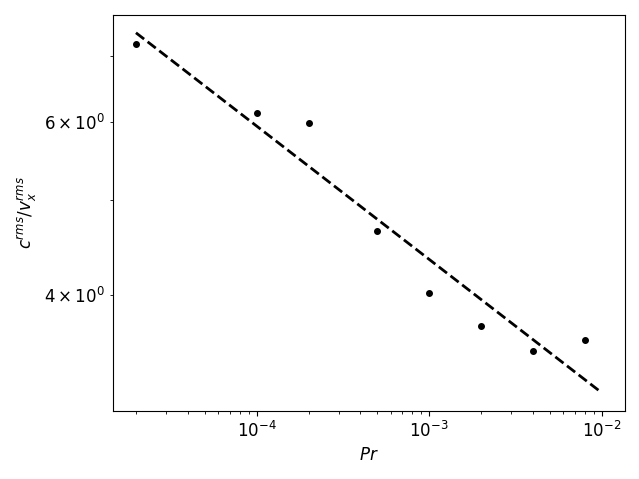} 
   \caption{Left: Scaling of $c^{rms}$ and $v_x^{rms}$ as a function of $Pr$. Scaling for c is $-0.55$ and for $v_x$ is $-0.42$. Right: Scaling of the adimensioned ratio $c^{rms}/v_x^{rms}$ as a function of $Pr$, scaling exponent is $-0.13$.}
   \label{fig_cvrms_pr}
\end{figure*}

These figures illustrate the effects of stable stratification on the flow and chemical concentration. Figure \ref{fig_corvxx} shows the longitudinal autocorrelation function on $v_x$ defined by equation \ref{eq_corvxx}, as a function of $\delta_x$ (left) and a rescaled $\delta_x$ (right). Figure \ref{fig_cvrms_pr} shows the impact of stratification on $v^{rms}$ and $c^{rms}$ and the associated fitted scalings.

\section{Variation with $Ro$}
\label{sec_appendix_ro}

\begin{figure*}[h!]
\centering
  \includegraphics[width=\textwidth]{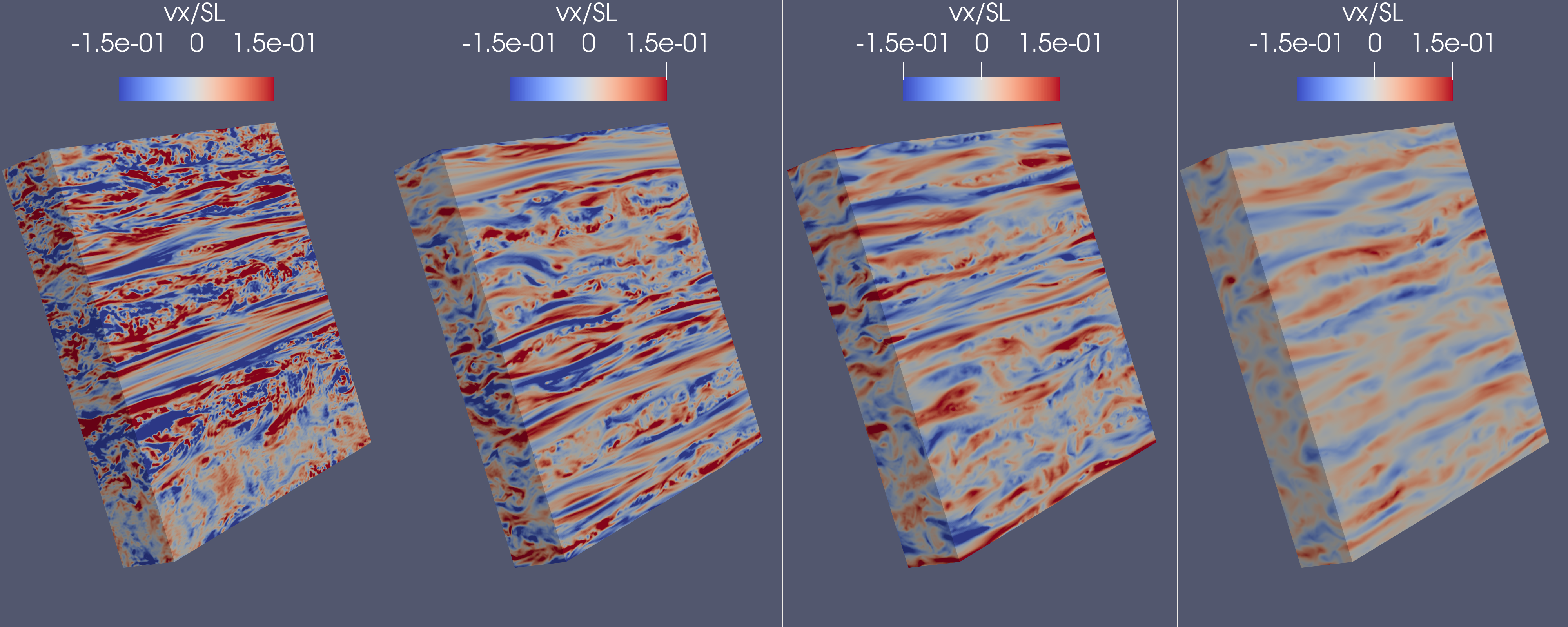}
   \includegraphics[width=\textwidth]{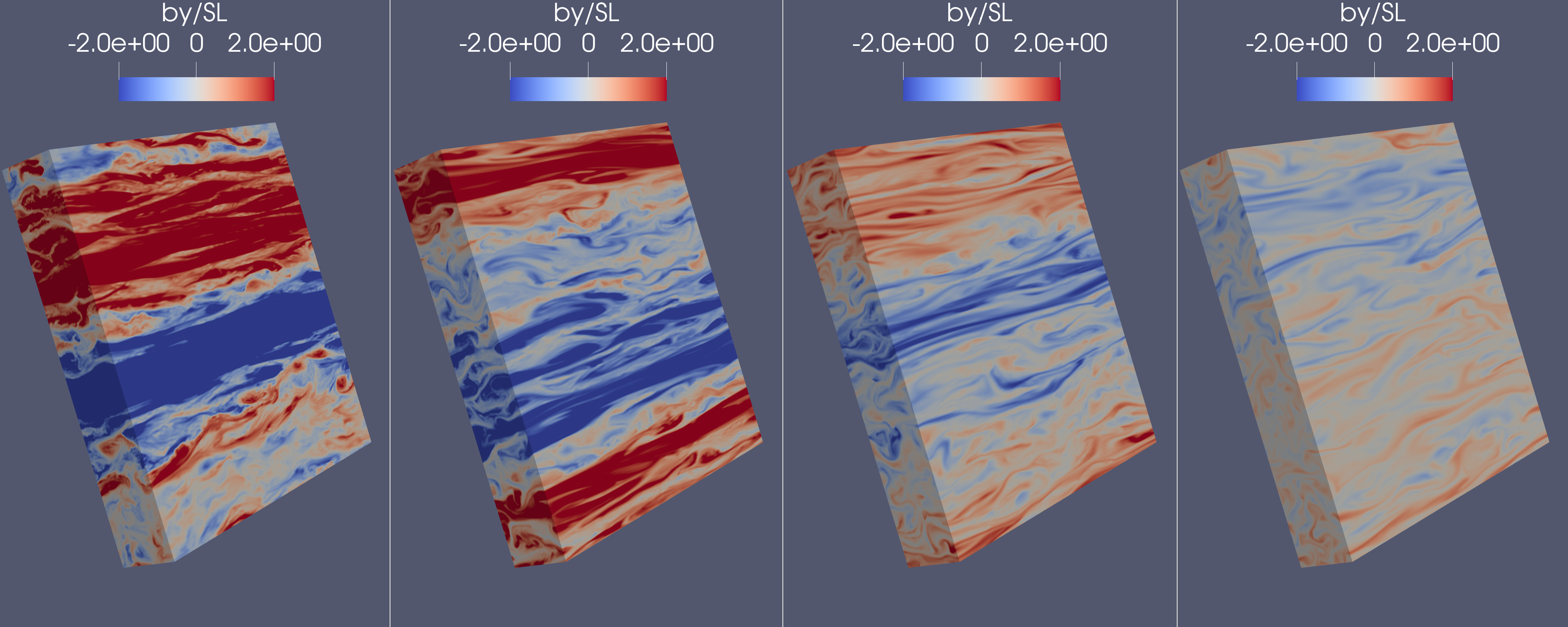}
   \caption{Snapshots of $v_x$ and $b_y$ as a function of the Rossby number (0.2, 0.5, 1 and 1.5 from left to right), with the same level of stratification $PrN^2/S^2=2.5\times 10^{-1}$. The impact of rotation is clearly visible particularly on the velocity field lengthscales and on the amplitude of the magnetic field.}
   \label{fig_snapshots2e-3}
\end{figure*}

\begin{figure*}[h!]
\centering
   \includegraphics[width=0.5\textwidth]{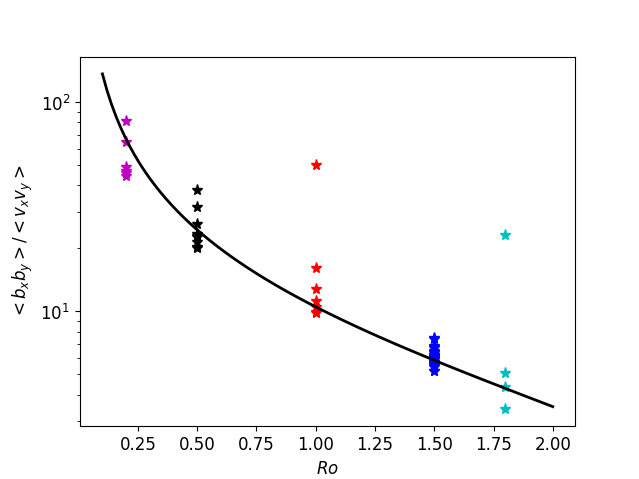}
   \caption{Ratio of Maxwell to Reynolds stresses as a function of the Rossby number, together with the black curve $y=3.5\times(4-Ro)/Ro$, based on the linear arguments of \cite{Pessah06}. The different points at the same $Ro$ correspond to calculations performed at different levels of stratification. The 2 outliers are cases at very high stratification ($PrN^2/\Omega^2>1$) where the Reynolds stress is almost $0$. }
   \label{fig_stressratio}
\end{figure*}

\begin{figure*}[h!]
\centering
\includegraphics[width=0.45\textwidth]{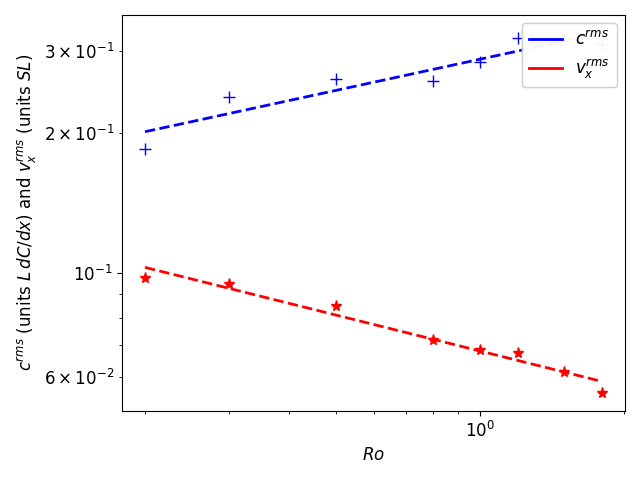} 
   \includegraphics[width=0.45\textwidth]{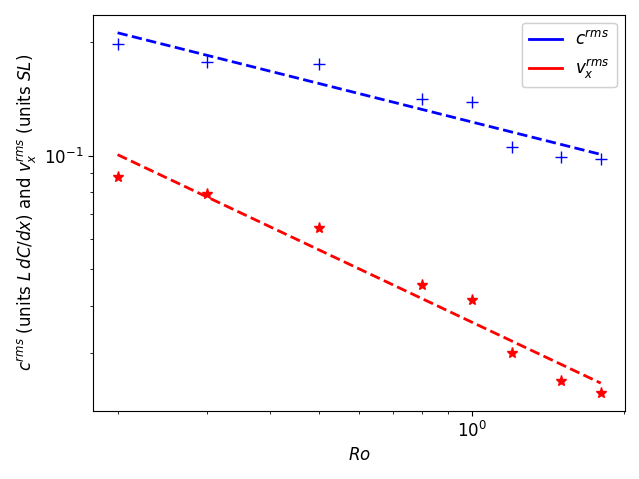} 
   \includegraphics[width=0.45\textwidth]{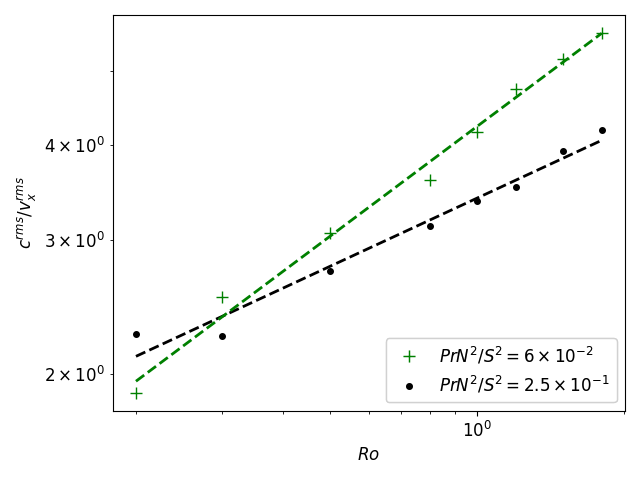} 
   \caption{Top panels: Scaling of $c^{rms}$ and $v_x^{rms}$ as a function of $Ro$ for a weak stratification $Pr=5\times10^{-4}$ (left) and high stratification $Pr=2\times10^{-3}$ (right). Scaling for c is $+0.22$ and $-0.25$ for $v_x$ for the weakly stratified case and $-0.34$ and for $v_x$ is $-0.64$ in the highly stratified case. Bottom panel: Scaling of the adimensioned ratio $c^{rms}/v_x^{rms}$ as a function of $Ro$ for the 2 levels of stratification, scaling exponents are $+0.48$ for the least stratified case and $+0.30$ for the most stratified one.}
   \label{fig_cvrms_ro}
\end{figure*}

These figures illustrate the impact of rotation on the flow, field and transport. Figure \ref{fig_snapshots2e-3} shows the $x$-component of the velocity and the $y$-component of the magnetic field as a function of the Rossby number, for a strongly stratified case at $PrN^2/S^2=2.5\times 10^{-1}$. Figure \ref{fig_stressratio} shows the ratio of the Maxwell to Reynolds stresses as a function of the Rossby number and the superimposed black curve $3.5 \times (4-Ro)/Ro$. This scaling $\propto (4-Ro)/Ro$ was proposed by \cite{Pessah06} based on linear arguments. The different points at the same $Ro$ correspond to calculations performed at different levels of stratification. Finally, figure \ref{fig_cvrms_ro} shows the dependencies of $c^{rms}$, $v_x^{rms}$ and $c^{rms}/v_x^{rms}$ as a function of Rossby, for two levels of stratification. This last quantity is thought to be a good estimate of the Lagrangian correlation time between the flow and the passive scalar $c$.

\section{Scaling exponents as a function of parameters}
\label{sec_param}

\begin{figure*}[h!]
\centering
   \includegraphics[width=0.49\textwidth]{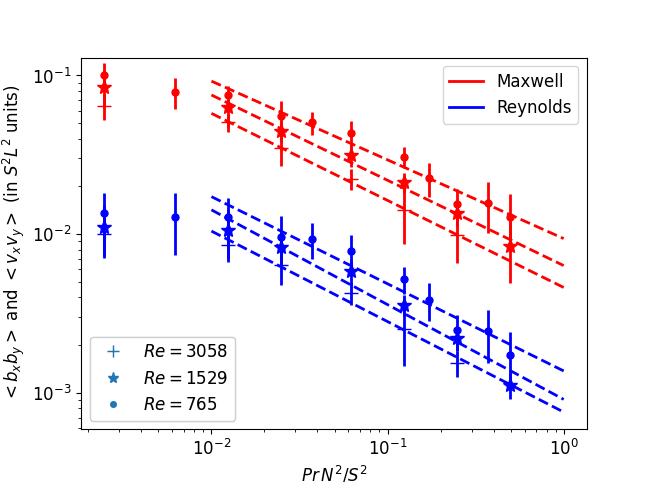}  
   \includegraphics[width=0.49\textwidth]{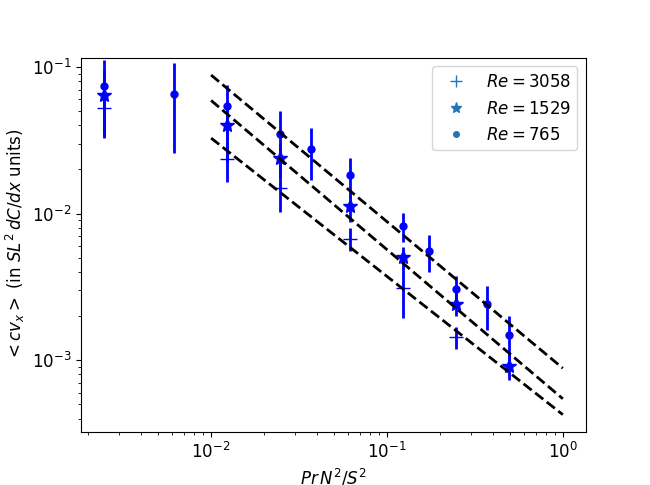}
   \caption{Stresses (left) and transport of chemicals (right) as a function of stratification for 3 different values of the Reynolds number, at $Ro=1.5$. Scaling laws are fitted independently for each Reynolds number. Values for the Maxwell stress are: -0.58, -0.56 and -0.56 and for chemicals: -1.05, -1.00, -1.02 for Reynolds numbers 765, 1529 and 3058 respectively.  }
   \label{fig_strat_param}
\end{figure*}

\begin{figure*}[h!]
\centering
   \includegraphics[width=0.49\textwidth]{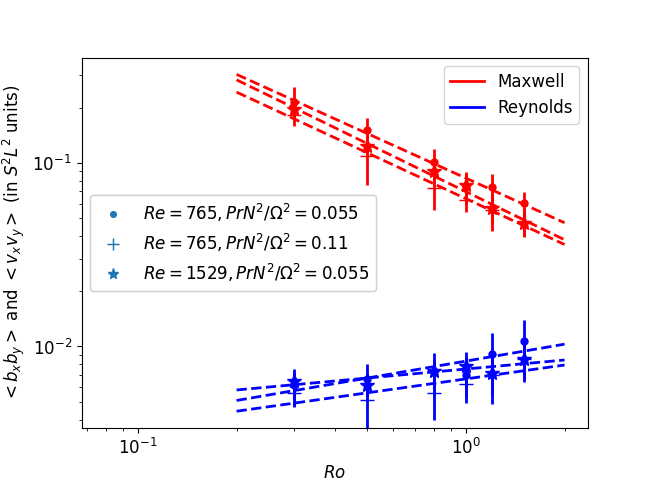}  
   \includegraphics[width=0.49\textwidth]{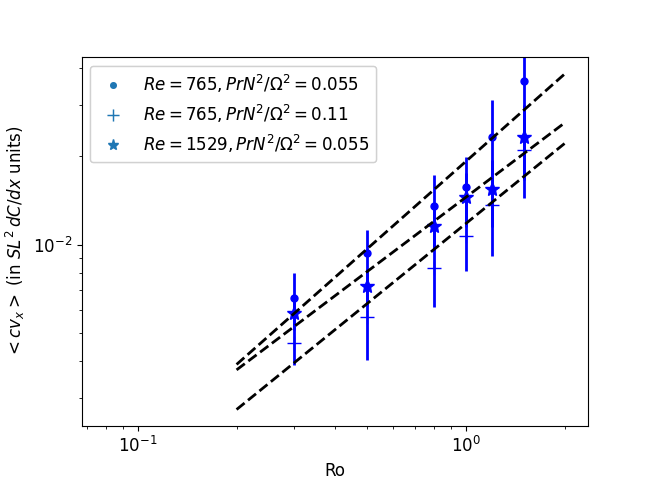}
   \caption{Stresses (left) and transport of chemicals (right) as a function of the Rossby number for 2 different values of the Reynolds number and 2 values of the stratification. Scaling laws are fitted independently for each set of parameters. Values for the maxwell stress are: -0.86, -0.83 and -0.87 and for chemicals: 0.92, 0.91, 0.84 for $Re=765, PrN^2/\Omega^2=5.5\times10^{-2}$, $Re=765, PrN^2/\Omega^2=1.1\times10^{-1}$ and $Re=1529, PrN^2/\Omega^2=5.5\times10^{-2}$ respectively.  }
   \label{fig_stress_ro_param}
\end{figure*}

We here show a study of how our scaling exponent in the Maxwell, Reynolds stresses and on the transport of chemicals depends on our various parameters, namely Reynolds number and stratification. Figure \ref{fig_strat_param} shows the different scalings obtained for the Maxwell and Reynolds stresses (left) and the flux of chemicals (right) as a function of stratification. Figure \ref{fig_stress_ro_param} shows the same quantities but as a function of the Rossby number.

\section{Variation with $Re$ and box size in the $x$-direction}
\label{sec_appendix_re}

\begin{figure*}[h!]
\centering
   \includegraphics[width=\textwidth]{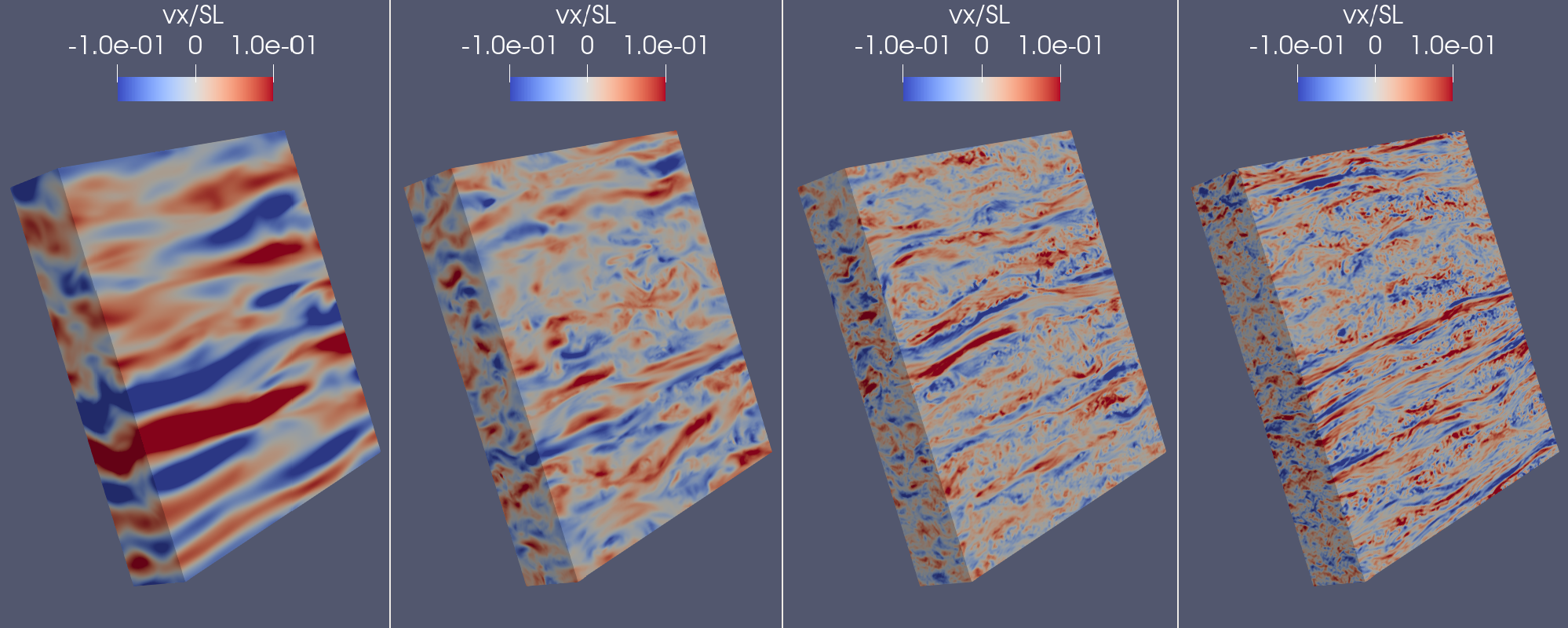}       
   \includegraphics[width=\textwidth]{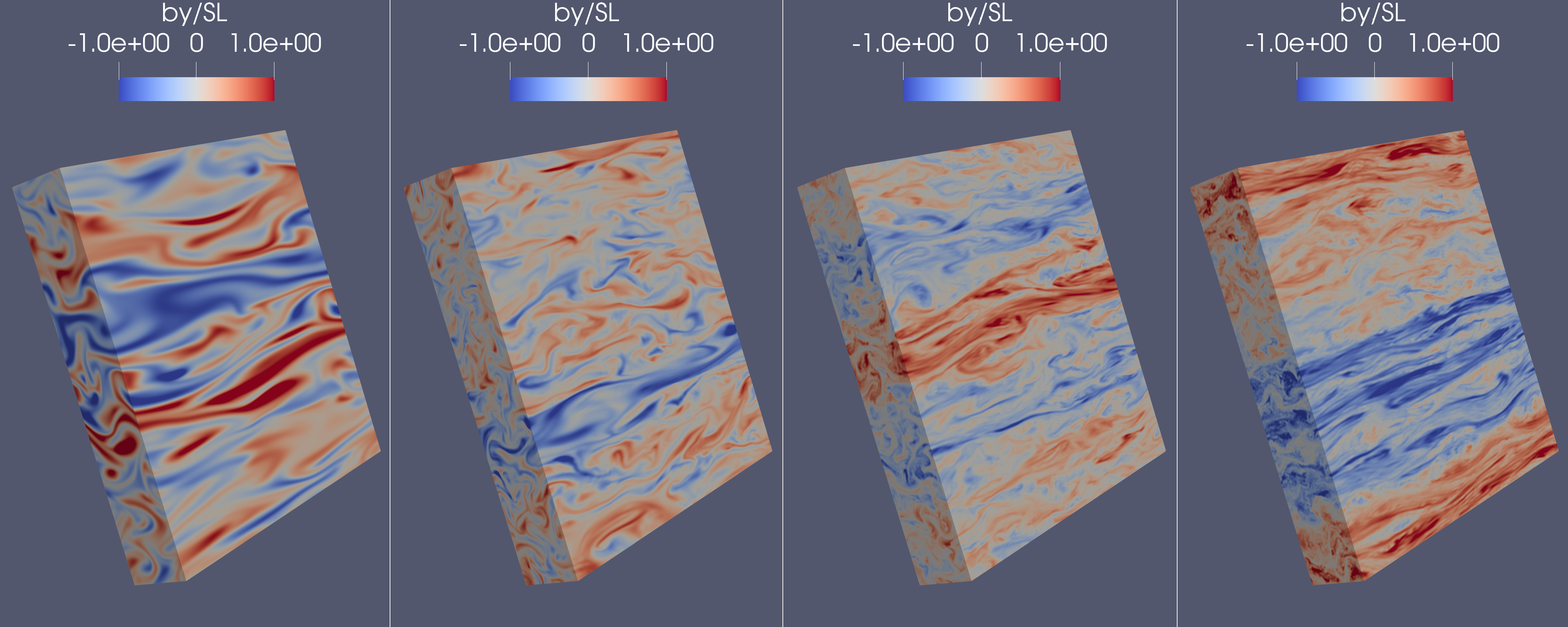}
   \caption{Snapshots of $v_x$ and $b_y$ as a function of the Reynolds number. Values of $Re$ are from left to right: 191, 765, 2294 and 4050. Stratification is fixed to $Pr N^2/\Omega^2=2.78\times 10^{-1}$.}
   \label{fig_refields}
\end{figure*}

\begin{figure*}[h!]
\centering
   \includegraphics[width=0.5\textwidth]{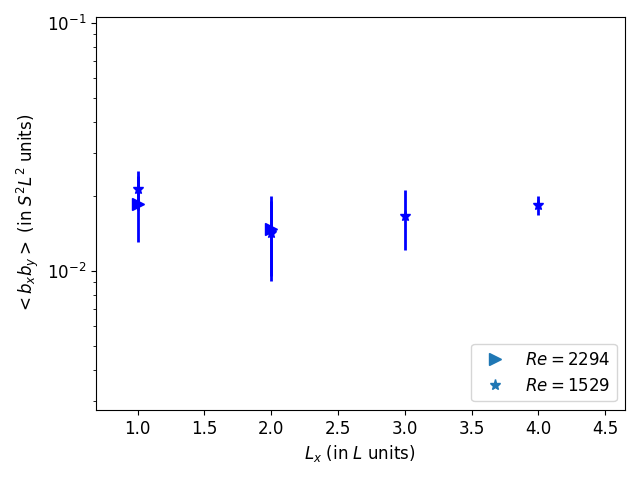}       
   \caption{Mean Maxwell stress as a function of $L_x$, with $L_y=L_z$ fixed. Stratification is fixed to $Pr N^2/\Omega^2=2.78\times 10^{-1}$ and $Re=1529$ and $2294$. The values of the stresses were computed only for $L_x=L$ and $L_x=2L$ for $Re=2294$.}
   \label{fig_stresslx}
\end{figure*}

\begin{figure*}[h!]
\centering
   \includegraphics[width=\textwidth]{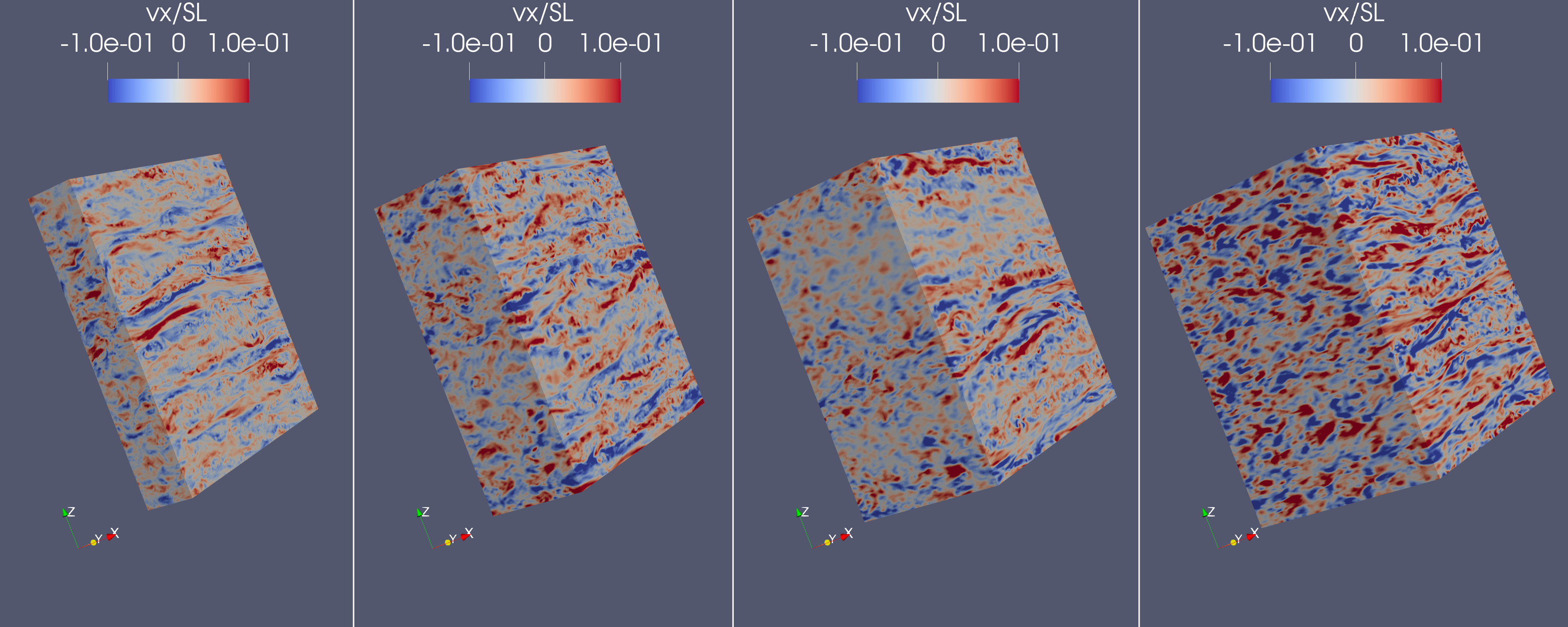}       
   \includegraphics[width=\textwidth]{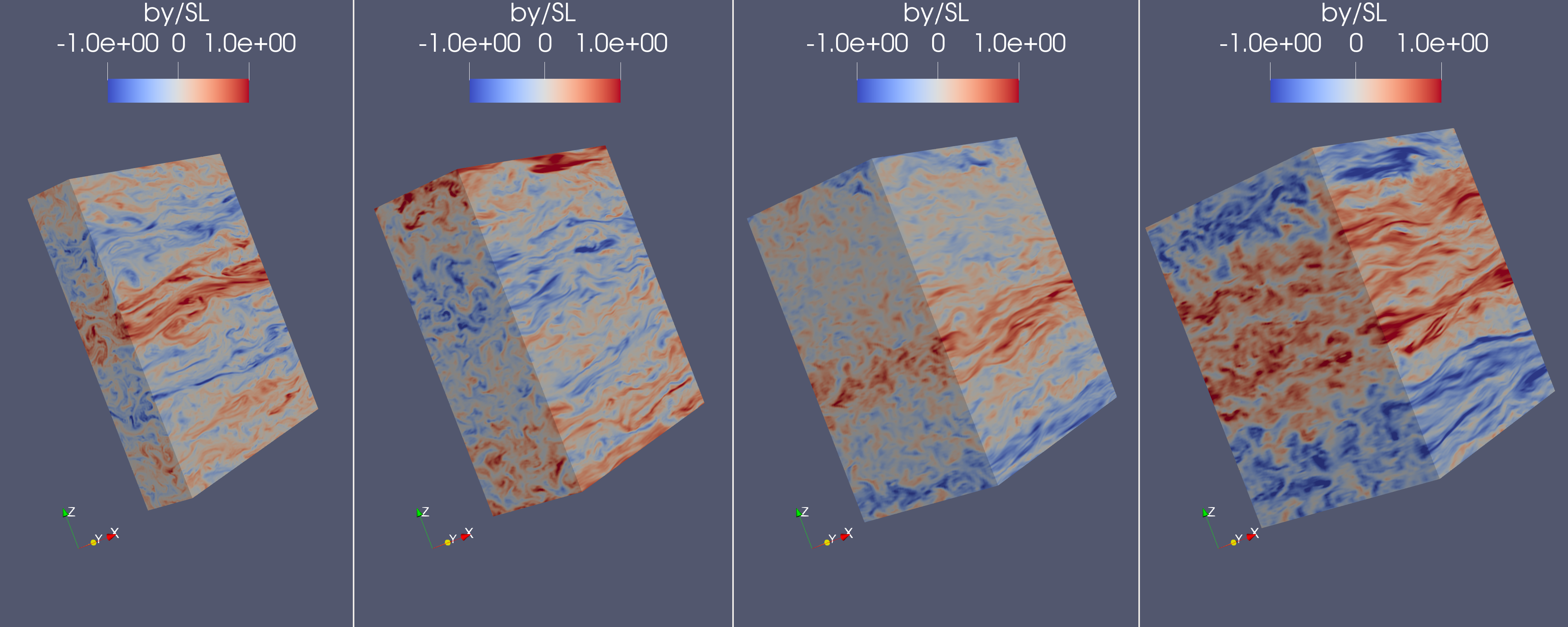}   
   \caption{Snapshots of $v_x$ and $b_y$ as a function of the box size in the $x$-direction. $L_x$ is varied from its initial value $L$ (fiducial run, left) to $2L$, $3L$ and $4L$. Stratification is fixed to $Pr N^2/\Omega^2=2.78\times 10^{-1}$ and the Reynolds number is chosen equal to $Re=1529$. Both the amplitudes and typical scales of both fields remain mostly unchanged when $L_x$ is varied at this sufficiently high values of stratification and Reynolds number.}
   \label{fig_fieldslx}
\end{figure*}

We here illustrate the effect of a varying Reynolds number and the box size $L_x$ in on our simulations. Figure \ref{fig_refields} shows the $v_x$ and $b_y$ components as a function of the Reynolds number $Re$, at a fixed level of stratification measured by $PrN^2/\Omega^2$. It is clear that at Reynolds numbers above $2\times 10^3$, the typical scales in the $x$-direction are much smaller than the box size and do not significantly vary when $Re$ is further increased. Figures \ref{fig_stresslx} and \ref{fig_fieldslx} illustrate the independence of our results (Maxwell stress in \ref{fig_stresslx} and $v_x$ and $b_y$ fields in \ref{fig_fieldslx}) on the box size in the $x$-direction for sufficiently high values of $Re$ and $PrN^2/\Omega^2$.

\end{appendix}

\end{document}